# Martian atmospheric disturbances from orbital images and surface pressure at Jezero Crater, Mars, during Martian Year 36

A. Sánchez-Lavega[1], E. Larsen[1], T. del Rio-Gaztelurrrutia[1], J. Hernández-Bernal[2], I. Ordóñez-Etxebarría[3], R. Hueso[1], B. Tanguy[4], M. Lemmon[5], M. de la Torre Juarez[6], G. M. Martínez[7], A. Munguira[1], J. A. Rodríguez-Manfredi[8], A.-M. Harri[9], J. Pla-García[8], D. Toledo[10], C. Newman[11]

[1] Escuela de Ingeniería de Bilbao, Universidad País Vasco, UPV/EHU, Bilbao, Spain (e-mail lead author: agustin.sanchez@ehu.eus)
[2] Laboratoire de Météorologie Dynamique, Sorbonne Université, Paris, France
[3] Planetario de Pamplona, Pamplona, Spain
[4] LESIA, Observatoire de Paris, Meudon, France.
[5] Space Science Institute, College Station, TX, USA.
[6] Jet Propulsion Laboratory/California Institute of Technology, Pasadena, CA, USA
[7] Lunar and Planetary Institute, Houston, TX, USA
[8] Centro de Astrobiología (INTA-CSIC), Madrid, Spain
[9] Finnish Meteorological Institute, Helsinki, Finland
[10] Instituto Nacional de Técnica Aeroespacial, INTA, Madrid, Spain
[11] Aeolis Research, Chandler, AZ, USA

## Abstract

We present a study of atmospheric disturbances at Jezero Crater, Mars, using ground-based measurements of surface pressure by the Perseverance rover in combination with orbital images from the Mars Express and Mars Reconnaissance Orbiter missions. The study starts at $L_s \sim 13.3°$ in MY36 (March 6$^{th}$, 2021) and extends up to $L_s \sim 30.3°$ in MY37 (February 28$^{th}$, 2023). We focus on the characterization of the major atmospheric phenomena at synoptic and planetary-scales. These are the thermal tides (measured up to the sixth component), long-period pressure oscillations (periods > 1 sol), the Aphelion Cloud Belt, and the occasional development of regional dust storms over Jezero. We present the seasonal evolution of the amplitudes and phases of the thermal tides and their relation with the atmospheric dust content (optical depth). Three regional dust storms and one polar storm extending over Jezero produced an increase in the diurnal and semidiurnal amplitudes but resulted in inverse responses in their phases. We show that the primary regular wave activity is due to baroclinic disturbances with periods of 2-4 sols and amplitudes ~ 1-15 Pa increasing with dust content, in good agreement with theoretical predictions by model calculations. The spacecraft images show a number of arc-shaped, spiral and irregular cyclonic vortices, traced by dust and clouds at the edge of the North Polar Cap, that could be behind some of the pressure oscillations measured at Jezero.

## Plain Language Summary

We study atmospheric disturbances observed as clouds and dust in images obtained from orbiting spacecraft, with simultaneous pressure measurements on the surface of Mars by the rover Perseverance on Jezero crater at latitude 18° North. The analysis focuses on the northern hemisphere in Martian Year 36. We study the seasonal evolution of the amplitudes and phases of the thermal tides (atmospheric oscillations in temperature and pressure forced by solar heating) and their relation with the dust content of the atmosphere and the presence of clouds, including the different types of dust storms reaching the crater Jezero. The images show a number of arc-shaped, spiral, ring-like and irregular rotating vortices, traced by dust and clouds at the edge of the North Polar Cap that could be behind some of the effects observed in pressure at Jezero. We detect wave activity with periods of 2-4 sols (a Martian day) and amplitudes ~ 1-15 Pascals that increase with the dust content, in good agreement with theoretical predictions by General Circulation Models.

**Key points**

• We study the surface pressure and the thermal tides and waves, from the first Martian Year, obtained by the rover Perseverance at Jezero.

• We correlate pressure variability and oscillations with dust storms, clouds and baroclinic cyclones studied from images taken from orbit.

• We present the relationships between the surface pressure and tides up to six components, with the optical depth measured at Jezero.

**1. Introduction**

Mars has a rich and varied meteorology, strongly influenced by daily and seasonal insolation cycles and where rapid changes develop due to the low atmospheric mass and a short radiative time constant (Zurek, 2017). Observations of Martian atmospheric phenomena have been typically carried out with telescopes on Earth and in near-Earth space, with spacecraft orbiting the planet, and with landers and rovers on the surface (James et al. 2017). Studies of particular meteorological phenomena are usually performed focusing on just one of these datasets. But, they rarely focus on both spacecraft imaging and in situ observations. In this work we use daily images obtained in the visible-optical range with the Visual Monitoring Camera (VMC) onboard Mars Express (MEX) (Sánchez-Lavega et al., 2018a; Hernández-Bernal et al. 2024b) and with the MARCI (Mars Color Imager) instrument onboard Mars Reconnaissance Orbiter (MRO) (Bell III et al., 2009; Cantor et al., 2010). We combine them with simultaneous in situ measurements from the Mars Environment Dynamics Analyzer (MEDA) instrument onboard the rover Perseverance in operation in Jezero Crater since 18$^{th}$ February 2021 with landing at longitude 77.45°E and latitude 18.44°N (Rodríguez-Manfredi et al. 2022, 2023).

This research focuses on synoptic (or mid-scale) and planetary-scale phenomena as they evolve in temporal scales approximately ≥ 1 sol, specifically on wave disturbances, dust storms and thermal tides (see Sánchez-Lavega et al., 2024 for a classification of space and time scales). Pressure has been shown to have high sensitivity and reliability as a reference magnitude to track the mid- and large-scale dynamical processes in the atmosphere (Hess et al., 1977, 1980; Ryan and Henry, 1979; Wilson and Hamilton, 1996; Harri et al., 2014, 2024; Banfield et al., 2020; Martinez et al., 2017; Rodríguez-Manfredi et al., 2023; Sánchez-Lavega et al., 2023 (paper 1); Zurita-Zurita et al., 2022; Hernández-Bernal et al., 2024a).

This study covers one Martian Year (see Cantor et al., 2010 for the MY numeration), starting on Perseverance sol 16 ($L_s \sim 13.3°$ MY 36, 6 March 2021) and extending to sol 720 ($L_s \sim 30.3°$ MY 37, 28 February 2023). During this period, pressure data were also acquired simultaneously by other surface missions located in nearly equatorial latitudes, the rover Curiosity (4.6°S, 137.4°E), the platform Insight (4.5°N, 135.6°E) that operated until 21 December 2022, and the rover Zhurong (25.07°N, 109.92°E) that was active between 27 June 2021 and 6 May 2022. Therefore, our analysis of the atmospheric perturbations detected in the images obtained from orbit can be of interest in further comparative studies of the pressure measurements performed at these other locations.

This article is organized as follows. In section 2 we present the data sources used and the methodology followed in the analysis of the surface pressure and orbital images. In section 3, we present the study of the pressure measurements in Jezero with sections 4 and 5 devoted to the study of thermal tides and baroclinic waves, respectively. In section 6 we present the different phenomena observed in the images of the northern hemisphere, ordered by solar longitude periods in which each atmospheric phenomenon develops. Finally, in section 7, we interpret these observations in the context of thermal tides, dust storms and baroclinic cyclones and their relation to the amount of atmospheric dust measured over Jezero. We conclude with a point-by-point listing of the main results.

**2. Data Analysis**

The methodology and techniques employed in this article for the analysis of MEDA pressure measurements have been described in previous articles (Rodríguez-Manfredi et al., 2023; Sánchez-Lavega et al., 2023, hereinafter called paper 1; Jaakonaho et al., 2023; Harri et al., 2024). In particular, the method to calculate the tidal components and long period waves was described in Sánchez-Lavega et al., (2023), which covered half of the period studied in this article (up to sol 460, $L_s \sim 241°$). Data are available in Rodriguez-Manfredi and de la Torre Juarez (2021).

For the survey of atmospheric phenomena, we have used images taken by two kind of cameras onboard two spacecraft with different and complementary orbits. MEX is a spacecraft in a polar orbit with pericenter at an altitude $\sim 300$ km, apocenter at an altitude $\sim 10,000$ km and an orbital period $\sim 7.5$ hr, allowing to observe Mars with different phase angles and local times (Sánchez-Lavega et al., 2024; Hernández-Bernal et al., 2024b). In contrast, MRO has a sun-synchronized nearly circular orbit, with an altitude over the surface of $\sim 250$ to 316 km and a period $\sim 1.86$ hr (Zurek and Smrekar, 2007). MRO observes the same area nearly every 24 hr at the same Local Martian Time, around 14-15 hr.

VMC is a frame camera of 640x480 pixels covering the spectral range $\sim 400-650$ nm with a Field of View (FOV) of 40°x30° (Ormston et al. 2011; Sánchez-Lavega et al. 2018a; Hernández-Bernal et al. 2024b). The camera takes a sequence of images in each orbital

4observation block assigned to the camera with a spatial resolution at nadir varying between ∼ 300 m and 12 km, the later allowing coverage of the entire Martian disk. We have analyzed VMC images (2024) using the software Elkano described in Hernández-Bernal et al. (2021). We have also analyzed the optical channel images of MARCI camera which has a wavelength coverage from 437 to 718 nm reaching a resolution of 1 km/pixel (Bell III et al., 2009; Cantor et al., 2010). MARCI captures observations of the entire planet every sol by combining data from 13 consecutive orbits (MARCI images, 2024). The Local True Solar Times (LTST) at the center of the observed swath range around 15:00 ± 02:00 hours. We combined MARCI frames (scans over the entire planet or tiles) and projected them in polar or rectangular maps in the *Meteomars* software (Ordóñez-Etxeberria et al. 2022). This software allows navigation (longitude and latitude grid and local time) and measurement of distances in the Martian surface. Complementarily, we use the software ISIS-USGS (2024) for the processing and map projection of selected MARCI tiles and QGIS (2024) for the corresponding measurements.

Because of the northern equatorial latitude of Perseverance at 18.5°N, we limit our image survey and analysis on synoptic-scale disturbances evolving in the northern hemisphere, with the exception of those phenomena that extend to both hemispheres, such as regional dust storms. A classification scheme of the studied phenomena in terms of either their temporal and spatial scales, or their location on Mars and seasonal distribution has been presented in figures 1-3 in Sánchez-Lavega et al. (2024). We have performed velocity measurements of global motions of weather systems and local motions of clouds and dust masses within them, following the identification and tracking of well-defined features (their centers or their edges) on images separated by a known temporal interval. In VMC, there is a wide variety of time intervals between images used for tracking. Typically, we have observing blocks separated by 30 minutes, 2-3 hours, and a whole day. In MARCI the time interval between successive swaths capturing the same region depends on latitude. Completing a polar orbit in about 2 hours, the spacecraft can observe high-latitude areas on every pass, providing higher temporal resolution. Near the equator, however, Mars's rotation shifts the location longitudinally, resulting in intervals of over 24 hours before the same region is revisited.

## 3. Surface pressure at Jezero

In Figure 1a we show the pressure measurements along each sol over the studied period and in Figure 1b the seasonal evolution (sol to sol) of the mean daily pressure in Jezero. The observed behaviour of the pressure follows the expected pattern from the cycle of polar condensation and sublimation of $CO_2$ and the influence of the amount of dust suspended in the atmosphere, measured as opacity in Fig. 1d, changes the daily range of pressure data (Fig. 1c) similarly to what was observed by Viking (Ryan and Herry, 1979). Lower dust content and higher cloud abundance (not shown, e. g. Fig. 4 in Smith et al., 2023) occurred approximately from $L_s \sim 0°$ to $L_s \sim 130°$, when sol-to-sol variations remained approximately constant. This period was followed by the epoch of high atmospheric dust opacity from $L_s \sim 130°$ to $L_s \sim 360°$ with larger sol-to-sol variations (Figure 1c). The arrival of dust storms over Jezero during this second epoch is revealed in the optical opacity measurements performed by Perseverance (Figure 1d). The optical depth was measured on a daily basis at different LMSTs from images obtained by the cameras Mastcam-Z and Skycam (Lemmon et al., 2022b) and TIRS on MEDA (Smith et al., 2023). The first regional storm over Jezero occurred at $L_s \sim$ 153-156° (sols 312-318) (labelled DS-1 meaning dust storm 1 in figures in this paper) and was analysed in previous papers (Lemmon et al., 2022; Sánchez-Lavega et al., 2023; Munguira et al., 2023). We extend its study in section 6.4. A north polar storm, not directly arriving to Jezero, is analysed in section 6.5. The second and third storms directly affecting Jezero corresponded to the regional events known as A ($L_s \sim$ 218°, sol 425 and labelled DS-A in figures in this paper) and C ($L_s \sim$ 314°, sol 575 and labelled DS-C in figures in this paper) (Kass et al., 2016; Martin-Rubio et al., 2024), and are studied in sections 6.6 and 6.7. A later increase of opacity corresponds to the close passage of a storm evolving at the northern edge of the polar cap ($L_s \sim$ 358°, sol 656; labelled as DS-NPC in figures in this paper) and is documented in section 6.8.

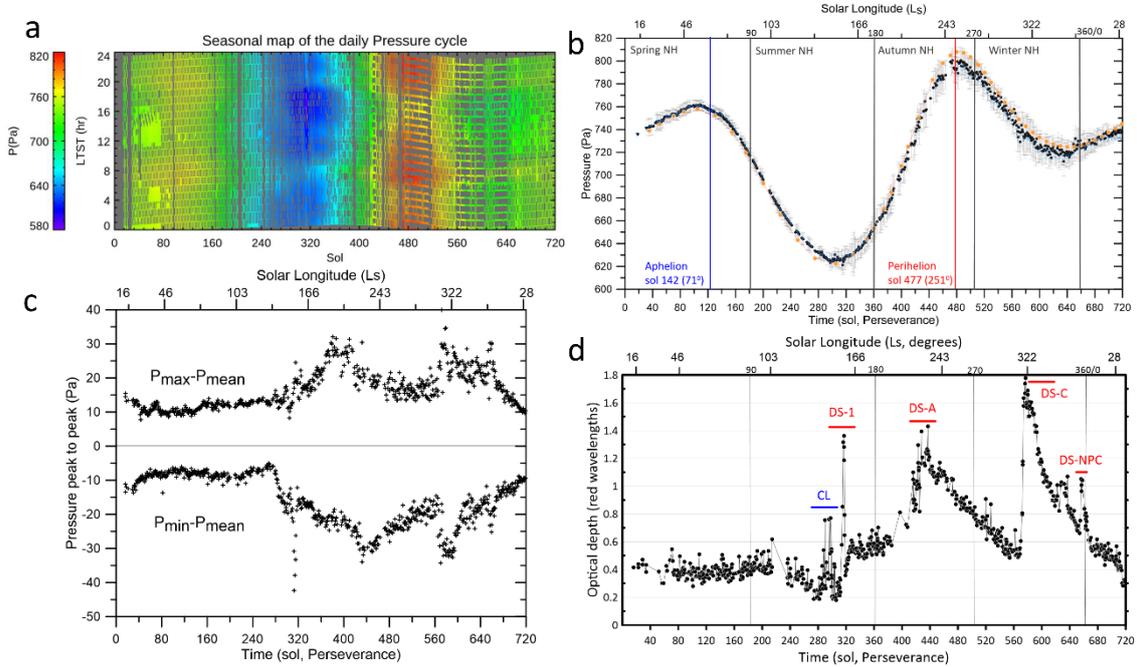

**Figure 1.** *Surface pressure and optical depth measured at Jezero. (a) Daily surface pressure (vertical axis LTST, Local True Solar Time) as a function of the sol number measured by MEDA on board Perseverance. The pressure is given in Pascal (Pa). Gaps in the measurements caused by limited resources of the mission and operations such as rock sampling appear in grey. (b) Seasonal evolution of the daily mean pressure between sols 16 and 716 ($L_s$ = 13° in MY 36 to $L_s$ = 30° in MY 37). Black dots and greys bars represent the mean pressure and its standard deviation for each sol. The orange dots are predictions by the Mars Climate Database for the standard climatology and average solar conditions. (c) Differences of maximum and minimum pressures relative to their mean value for each sol; (d) The seasonal evolution of the visible-optical depth measured with Skycam (600-800 nm) and Mastcam-Z (880 nm) on board Perseverance. The identifications correspond to the following cases: CL (clouds), DS-1 (Dust Storm 1, first regional storm over Jezero), DS-A (Dust Storm A), DS-C (Dust Storm C), DS-NPC (Dust Storm at the edge of the North Polar Cap passing close to Perseverance).*

## 4. Tide amplitudes and phases

Figures 2a-f show the seasonal evolution of the amplitude of the six tidal components ($S_i$, i =1-6) corresponding to periods of 24, 12, 8, 6, 4.8 and 4 hr (see Paper 1 for details of the measurements and data analysis). For $L_s$ >130°, the correlation of the amplitudes of the diurnal ($S_1$) and semidiurnal ($S_2$) components with the optical depth is evident because both follow the dust storm events (Fig. 2a-b; see section 7.1). The amplitudes $S_3$ to $S_6$ follows a dual behavior. The odd components $S_3$ and $S_5$ show a similar pattern with a peak amplitude on sols ~ 90 - 100 ($L_s$ ~ 50°) and a minimum around sol 200($L_s$ ~ 100°) (Fig. 2c-e).The even components $S_4$ and $S_6$ exhibited nearly simultaneous maximum amplitudes at sols ~ 380 ($L_s$ ~ 191°) and ~ 640 ($L_s$ ~ 350°) and a broad minimum from sol ~ 80 to 260 (Fig. 2d-f). This dual behavior was also observed by Insight for $S_3$ to $S_6$ (Hernández-Bernal et al., 2024a). As a comparison, we show the calculations according to the Mars Climate Database (MCD v 6.1) by the Laboratoire de Météorology Dynamique (LMD) (Forget et al., 1999) for standard climatology and average solar conditions, and the pre-landing predictions from the Mars Weather Research and Forecasting (MarsWRF) for the innermost nest for Jezero crater (domain 5 with a horizontal resolution of about 1.5 km) (Newman et al., 2021). In general the Martian PCM results reproduce the behavior of the tidal amplitudes reasonably well. There is some disagreement in $S_1$ in the range of sols 120-290 ($L_s$ = 62°-141°) and for $S_3$ around sols 40-160 ($L_s$ = 25°-80°), but in general the agreement is very good, in particular for $S_2$ and $S_4$-$S_6$, which validates the measurements. A detailed study of the relationship between tides and water ice and dust optical depth is presented in section 7.1.

The seasonal behavior of the phases of the six tidal components is more complex (Figure 2 g-h). Components $S_1$-$S_3$ showed varying degrees of variability up to sol 290 ($L_s$ ~ 140°). The phase of the diurnal component showed a change of ~ 11 hr (~ ½ cycle) between sols 130 and 210 ($L_s$ ~ 66°-102°), and then had punctual decreases of 2-4 hr during the storms except for storm DS-C that underwent a complete phase reversal (24 hr). The phase of the semidiurnal component showed first a large drop of 4 hr with a minimum in sol 280 ($L_s$ ~ 136°) and then punctual increases of ~ 2 hr during the four storms. The phase of component 4 is relatively constant, while the phase of component 3 was noisier and showed a larger variability. The phases of tides 5 and 6 (not shown in Figure 2) were significantly noisier during that period and could be affected by the measurement strategy of the MEDA instrument resulting in many gaps in MEDA measurements (grey regions in Figure 1a; for a description of the MEDA measurement strategy see Rodríguez-Manfredi et al. 2023). In particular, the dusty period was abundant in rock sampling activities that further limited MEDA measurements with many sols with pressure data acquired in blocks of 15 min separated by almost one hour.

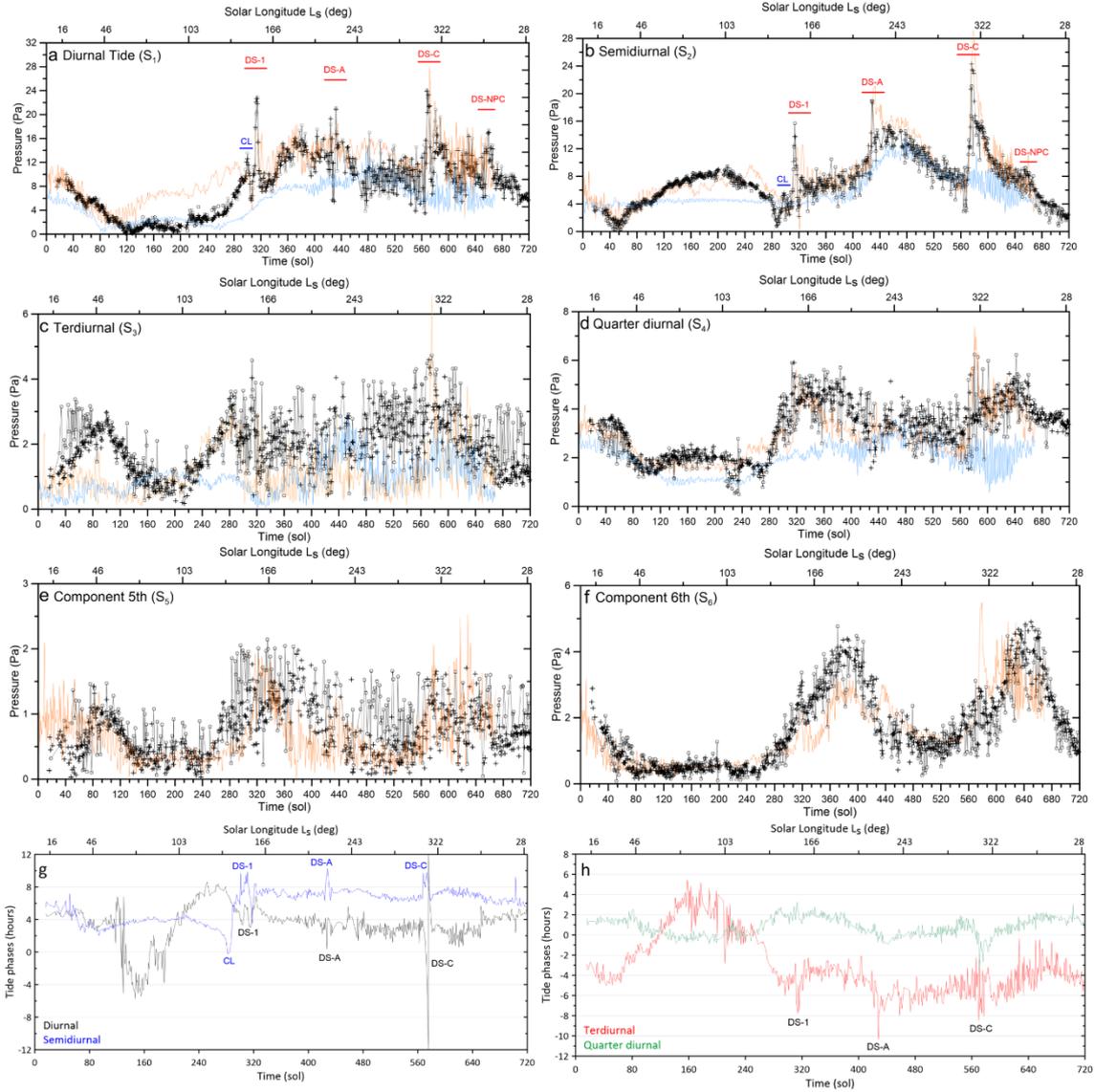

**Figure 2**. *Seasonal behaviour of the amplitudes and phases of the tidal components. (a)-(b) Diurnal and semidiurnal amplitudes $S_1$ and $S_2$ (black points), with clouds (CL) and dust storm events (DS) identified as in Figure 1. (c)-(f) Amplitude of the components 3 to 6 ($S_3$ to $S_6$) (black points). In (a-f) the MCD predictions are shown by the orange line and in (a-d) the GCM-MarsWRF model predictions by a blue line. (g) Phase of the diurnal (black line) and semidiurnal (blue line) components with features identified as in (a). (h) Phase of the terdiurnal (red line) and quarter diurnal (green line) components with features identified as in (a).*

## 5. Long-period waves

We have studied the long-period oscillations (> 1 sol) that occur around the daily mean pressure trend, using the same methodology as in paper 1, i. e. calculating the residuals to polynomial fits for sol sectors that typically cover > 60 sols according to the pressure cycle shown in Fig. 1b. The resulting detrended pressure is shown in Figure 3a, where the large-scale atmospheric phenomena are labelled in the figure. The results clearly show the existence of two epochs. In the first epoch, corresponding approximately to the first half of the Martian year with less suspended dust (Fig. 1d), the peak to peak amplitudes of the long-period waves are low, on average 1.6 Pa (sols 80-288, $L_s \sim 40°$-140°). In the second epoch, corresponding to the dusty epoch (sols 288-720, $L_s \sim 140°$-30°), the amplitudes are higher, on average ~ 4.2 Pa, but with peaks reaching up to 18 Pa. The most pronounced increases in amplitude (up to ~ 5-8 Pa larger) took place during dust storms (DS-1, DS-A, DS-C, DS-NPC). We also made a tentative identification of the impact of other phenomena, such as the dust spirals that develop at North Polar Cap edge (producing amplitudes ~ 2.5 Pa) and the Double Annular Cyclone DAC (~ 1.2 Pa). The details for the different cases studied in the images are presented in the following sections.

We have calculated the period of these oscillations as in paper 1 from the time separation between consecutive peaks. The result is shown in Fig. 3b. The dominant period is between 2 and 4 sols, with a mean value of 3.8 ± 1.9 sol.

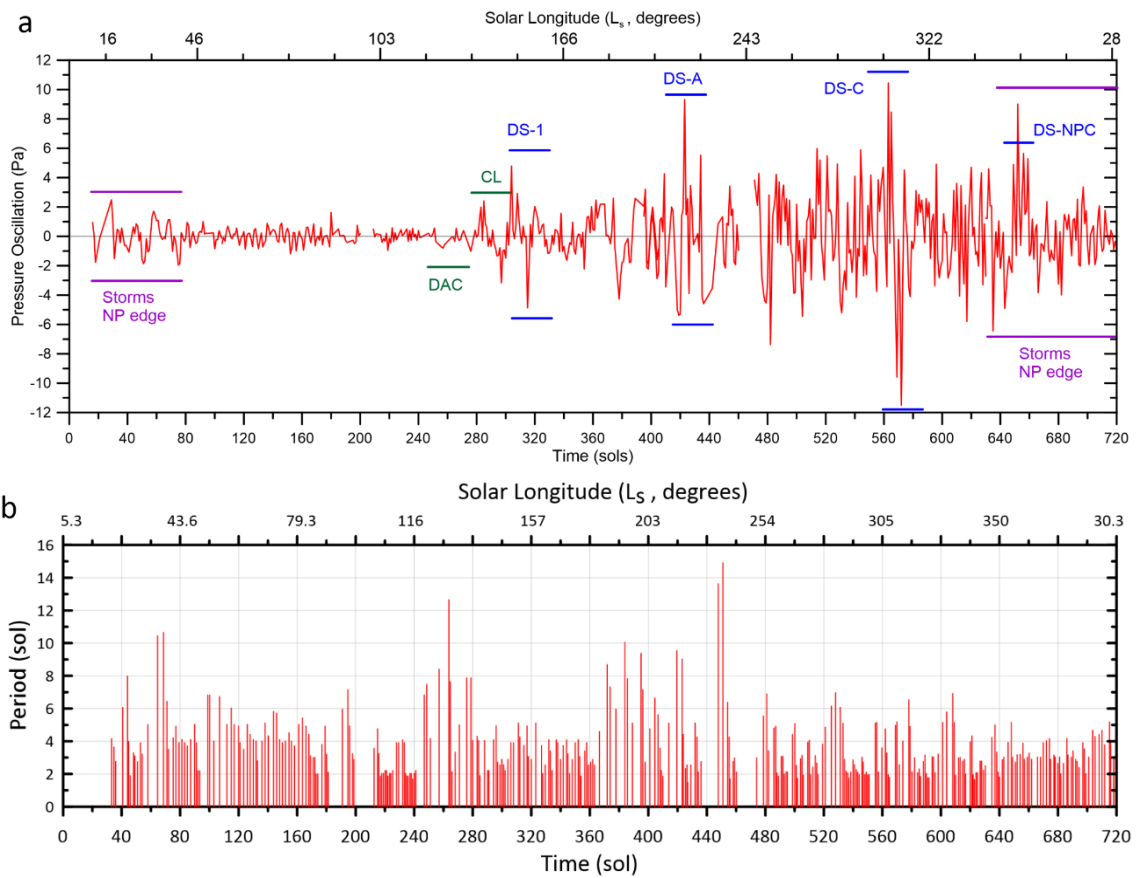

**Figure 3.** *Seasonal evolution of the long-period pressure oscillations (> 1 sol) at Jezero over one Martian Year. (a) Detrended pressure with some of the phenomena discussed in the text identified by their acronyms. Dust storm spirals at the edge of the North Pole Cap and the double annular cyclone (DAC) took place along the sols between magenta and green bars, respectively. The dust storms 1 (DS1), A (DS-A), C (DS-C) took place along the sols between blue bars. (b) Period of the oscillations.*

## 6. Imaging synoptic and planetary-scale disturbances

In this section we classify and study the different types of synoptic and planetary-scale atmospheric phenomena visible in VMC/MEX and MARCI/MRO images, and look for their possible relationship with the pressure observations previously described. We focus on disturbances in the Northern Hemisphere, but we also consider regional dust storms propagating from the South. Sánchez-Lavega et al. (2024) present a schematic classification of the main Martian atmospheric disturbances. In the specific case of the Dust Storms, an analysis of their statistics, size and areographical and temporal distribution during this same Martian year (MY 36) can be found in Guha et al. (2024).

### *6.1. North Pole edge dust cyclones* ($L_s = 0°-90°$)

Intense atmospheric activity occurs at the edge of the North Polar Cap during the spring season. Dust storms with a variety of shapes (irregular, textured, arcs and spirals) and sizes ranging from ~ 1000 to 2000 km form and disappear in short temporal scales of 1-3 sols (Hinson & Wang, 2010; Guzewich et al., 2015; Montabonne et al., 2015; Heavens, 2017; Barnes et al., 2017; Kahre et al., 2017; Sánchez-Lavega et al., 2018a, 2022, 2024) (Fig. 4). They are identified as "Storms NP edge" in the pressure amplitude oscillations (Fig. 3a). There can appear up to 4-6 cyclones per latitude band (Fig. 4b), with some long arcs reaching lengths of ~ 4000 km. In some cases there are water-ice cloud fields associated to these storms, (Fig. 4c-d), indicating that the disturbances extend from the surface (where they lift the dust) to the middle atmosphere where they form condensation clouds. Most intense activity takes place in the longitude sector 310°E-330°E and in latitudes 50°N-80°N, following the retreat of the North Polar Cap (Fig. 5). These storms move predominantly eastward, sometimes with a northward component (direction indicated by arrows in Fig. 5). We have measured averaged translation velocities $<V> = +20$ to $+35$ ms$^{-1}$ for the arc-shaped features (VMC/MEX images, sols 16-38, $L_s = 14°-25°$, March 2021) and $<V> = 3$ to $16$ ms$^{-1}$ for ten spiral cyclones (MARCI/MRO images between sols 40 and 800($L_s = 25°-43°$, April-May 2021). Taking into account that the time intervals are longer than one day and that the resolution of MARCI images is ~ 1 km (nadir view) and of VMC ~ 5-10 km (average MEX distance to Mars), the errors in the velocity measurements are very low. However we estimate that intrinsic changes in the shape of the tracers and in the cursor pointing can give velocity errors of 1-2 ms$^{-1}$. The shortest distance to Perseverance reached by one of these spirals was ~ 2,600 km in sol 40 ($L_s \sim 25°$, 31 March 2021).

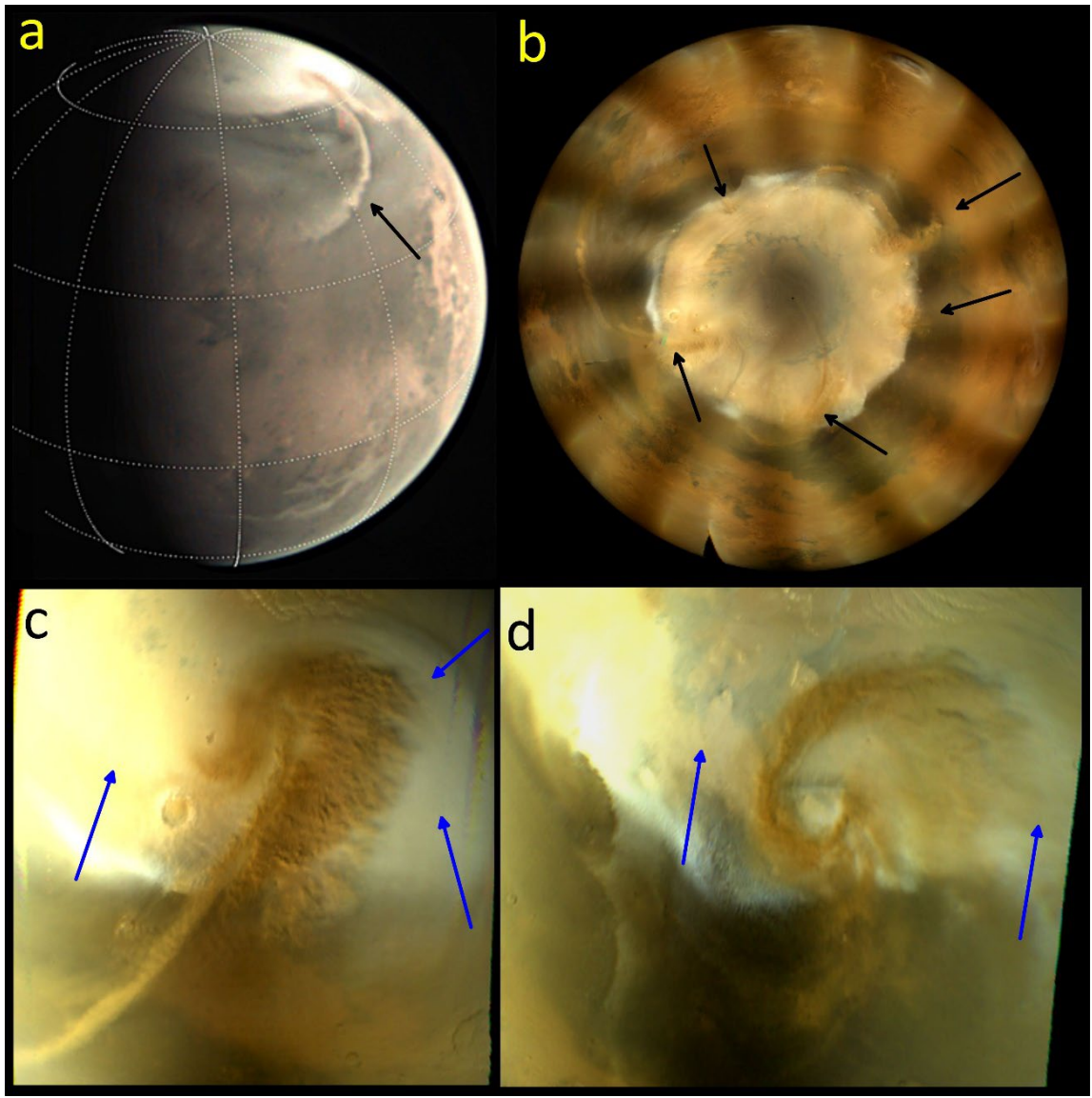

**Figure 4.** *Selected cyclone Dust Storms at North Polar Cap edge ($L_s$= 0°-45°) marked with black arrows in (a)-(b). (a) 30 March 2021, sol = 39, $L_s$= 24.5° (VMC/MEX). (b) Polar map showing a series of five storms, 31 March 2021, sol = 40, Ls= 24.9° (MARCI/MRO). (c) 31 March 2021, sol = 40, $L_s$= 24.9° (MARCI/MRO). This is the same storm as in (a), and the leftmost in (b). (d) 17 May 2021, sol = 86, Ls = 46.4° (MARCI/MRO). The blue arrows in (c)-(d) show the accompanying water ice clouds.*

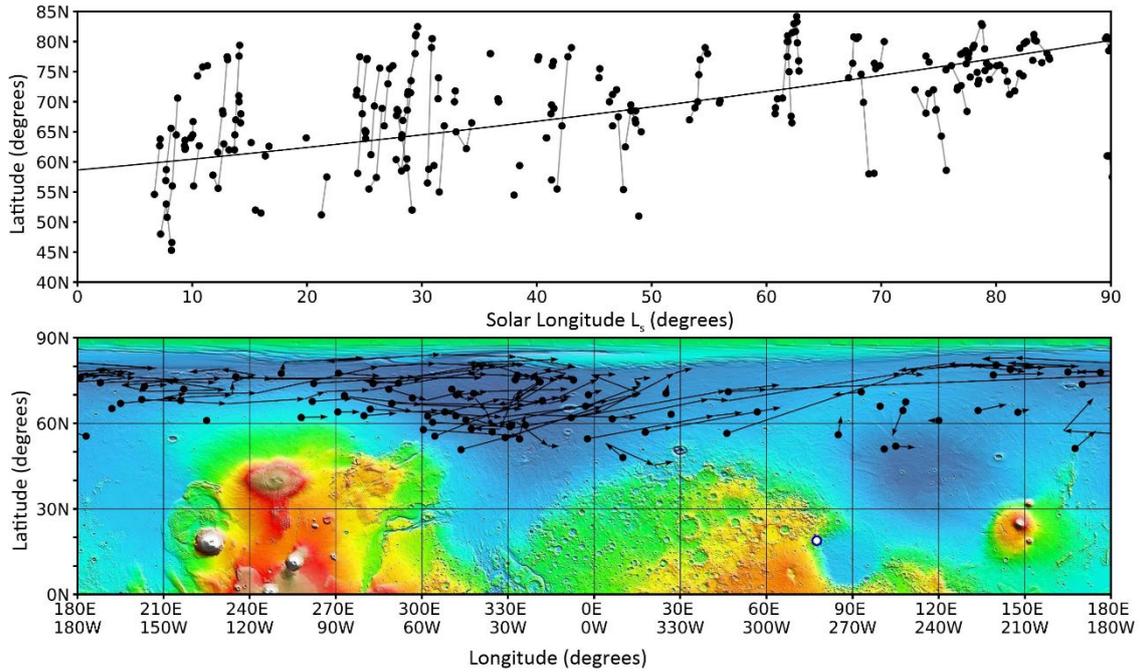

**Figure 5**. *(a) Latitude location and displacement of dust storms (arcs, spirals, irregular shape) at the edge of the North Polar Cap for $L_s$ = 0°-90° (dot: center of the feature). A same feature is identified by a black line joining consecutive dots. (b) Map showing the tracks (black line joining dots) of the storms shown in (a). The arrows mark the motion direction. The location of Perseverance is indicated by the blue-outlined white disk.*

### *6.2. Double-annular cyclone (DAC) ($L_s$ = 120°-140°)*

The DAC is a seasonally recurrent disturbance that forms at dawn in the longitude range ~ 270°-330°E and at subpolar latitudes near 60°N, dissipating during diurnal hours, and reappearing again the next sol (Gierasch et al., 1979; Cantor et al., 2002, 2010; Sánchez-Lavega et al., 2018b). Its vorticity is cyclonic with measured tangential velocities ~ 5-20 ms$^{-1}$ (± 1 ms$^{-1}$) at 10 km altitude. It appears sometimes as a single ring of water-ice clouds (marked as "center" in Figure 6) and sometimes with two coupled cyclones (east and west), each with a size of 600-800 km (Sánchez-Lavega et al., 2018b). During MY 36, it was observed from $L_s$ ~ 116° to 134° (sols 239-276) (Figure 6). The western cyclone was placed at latitudes 60°N to 66°N, slightly northward from the eastern cyclone at latitudes 58°N to 62°N, with both occupying a large zonal extent in longitude from ~ 255°E to 345°E. Their centers remained separated by about 1550 km. Both cyclones moved northeast with mean velocities <V> = +0.5 to +3.7 m/s (derived from tracking their centers on MARCI images). The shortest distance to Perseverance reached by the eastern cyclone was ~ 4,500 km in sol 278 ($L_s$ ~ 135°, 30 November 2021).

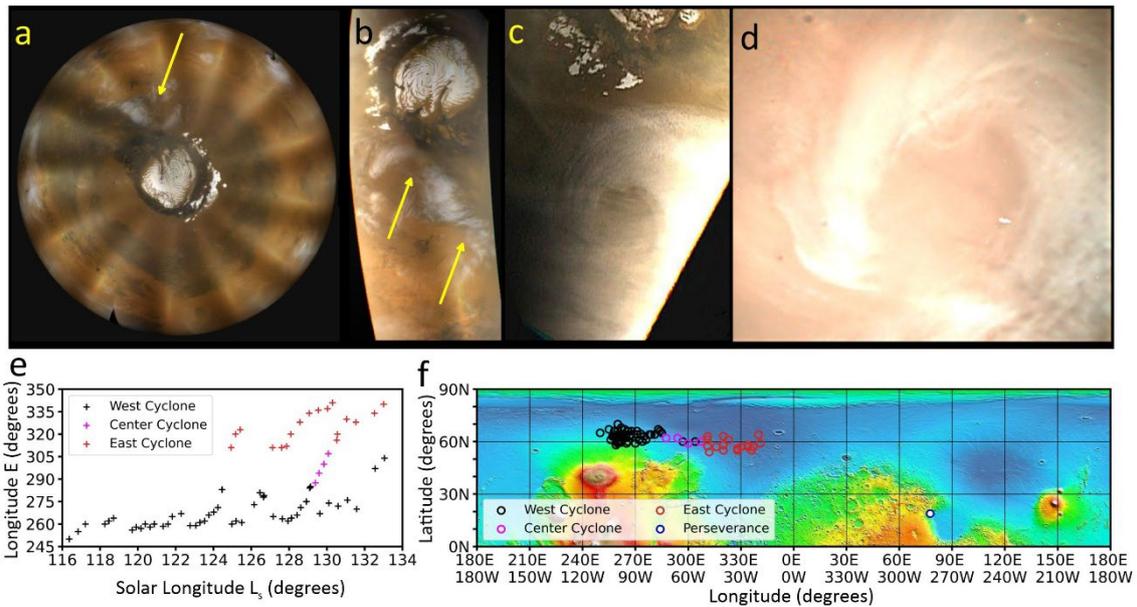

**Figure 6**. *The double annular cyclone (DAC). (a) and (b) Two views of DAC on 12 Nov. 2021 (sol 260, Ls = 126°) (MARCI). (c) DAC well developed on 29 Oct. 2021 (sol 247, Ls = 119.6°) (MARCI).. (d)DAC on 17 Nov. 2021 (sol 265, Ls = 128.6°) (VMC). (e) Plot showing the longitude location of the two DAC cyclones in time (West and East edges and Center). (f) Map showing the displacement in longitude and latitude of the two DAC components.*

### 6.3. Aphelion Cloud Belt ($L_s$ = 55°-140°)

The Aphelion Cloud Belt (ACB) develops progressively during the aphelion season, when dust content is minimum (Wang & Ingersoll, 2002; Clancy et al., 2017, Wolff et al., 2019). Water-ice clouds form over the volcanos and cover a great part of the equatorial band between latitudes ~ 30°N and 20°S, with a large cloud concentration from longitudes 220°E to 360°E (Fig. 7a).

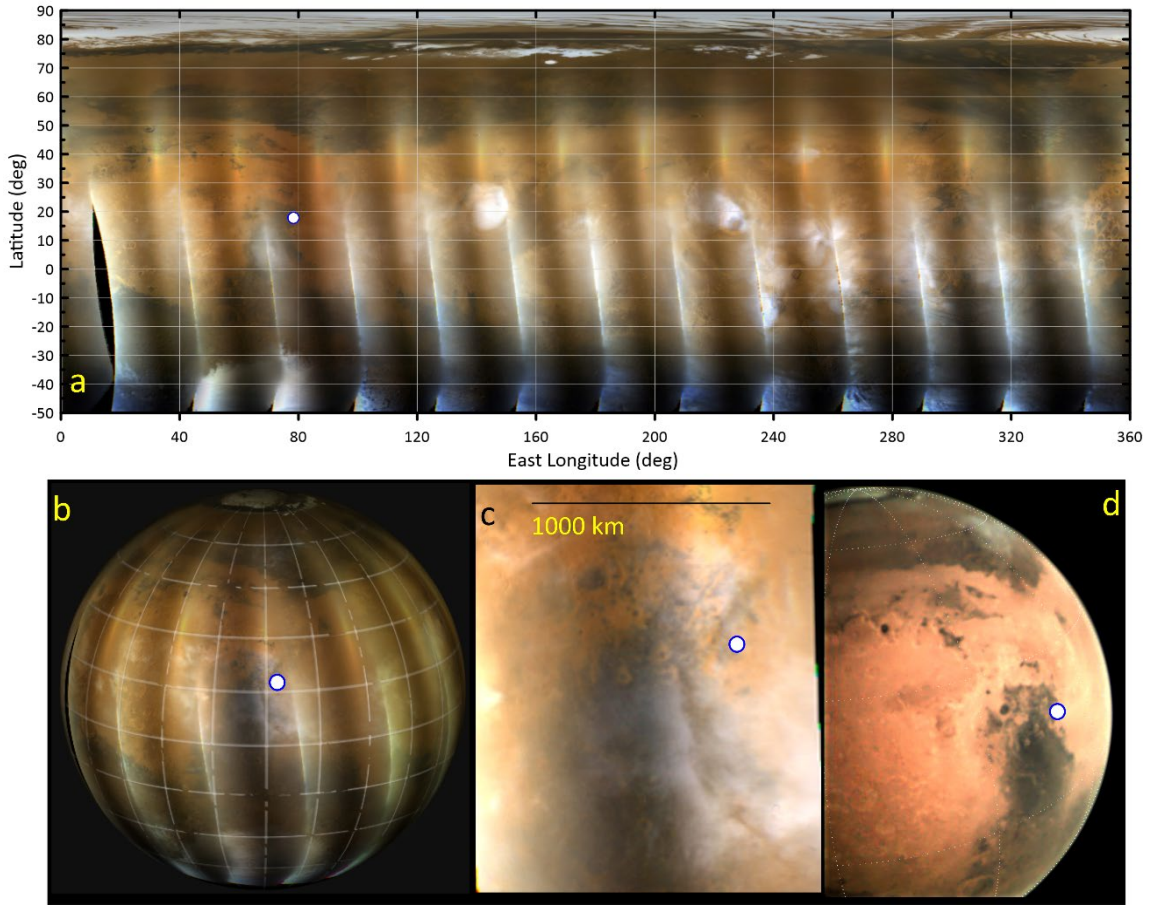

**Figure 7.** *(a) Rectangular map projection showing the Aphelion Cloud Belt on 30 September 2021 (sol 218, $L_s$= 106°). Clouds over Jezero: (b) MARCI/MRO image on 14 December 2021 ($L_s$ = 142°, sol 291, LTST 15.1 hr). (c) Enlargement of the previous image and of the Jezero area. Scale indicated by black line. (d) VMC/MEX, 29 December 2021 ($L_s$ = 149°, sol 306, LTST 11.2 hr). In all panels, the location of Perseverance is indicated by the blue-outlined white disk.*

Inspection of MARCI images during this period shows the central part of the Syrtis Major area covered with water-ice clouds towards $L_s \sim 100°$, with a higher density of clouds reaching Jezero mostly on sols 284-305 ($L_s \sim 138°$-149°)(Fig. 7b-d), increasing the optical depth of the atmosphere (Lemmon et al., 2022a; Smith et al., 2023). This is consistent with the high density cloud detection from $L_s \sim 130°$-150° by two different Perseverance instruments by Toledo et al. (2023) and Patel et al. (2023).

The development of the ACB and the arrival of clouds at Jezero affected the diurnal and semidiurnal tides (Fig. 2a-b, 2g), which followed a complementary trend. Hinson & Wilson (2004) have shown the strong coupling that exist between the thermal tides and the radiatively active water ice clouds. With the beginning of the ACB period, $S_1$ decreased, becoming essentially null in sols 117-131 ($L_s \sim 60°$-67°) and in sols 189-199 ($L_s \sim 93°$-97°). Then, with clouds present in Jezero, $S_1$ has a maximum in sol 300 ($L_s \sim 146°$) followed by a minimum in sol 306 ($L_s \sim 149°$), with a maximum-to-minimum pressure change of 7 Pa. This behaviour of $S_1$ follows the evolution of the cloud opacity measured by Toledo et al. (2023) and Patel et al (2023) that found an increase in cloud opacity around $L_s \sim 120°$-$150°$. The $S_2$ component had two deep minima reaching essentially zero at the beginning of the ACB period in sols 46-57 ($L_s \sim 28°$-$33°$) and at its end in sols 287-290 ($L_s \sim 139°$-$141°$), and was larger than the diurnal component during most of the ACB period.

The phase of the diurnal component showed a deep and prolonged decrease of about 8 hr (1/3$^{rd}$ of the daily cycle) from sols 130-210 ($L_s \sim 66°$-$102°$) and then increased to a maximum in sols 235-286 ($L_s \sim 114°$-$139°$). In parallel, the phase of the semidiurnal component showed a drop of 4 hr (1/3$^{th}$ of the semidiurnal cycle of 12 hr) starting in sol 230 ($L_s \sim 112°$) but with a deep minimum in sol 286 ($L_s \sim 139°$), just before the arrival of DS-1.

### *6.4. Regional Dust storm over Jezero ($L_s = 150°$ - $156°$)*

The first dust storm of the season in Jezero (DS-1) initiated on 1 January 2022 ($L_s = 150.7°$, sol 308) close to the edge of the south polar cap (latitudes 35°S to 45°S and 95°E to 135°E) (Fig. 8). The storm expanded rapidly northward, and the point closest to Jezero in VMC images at 22:38 UT (Fig. 8a), at 10°S, 100°E, was at about 3400 km from the rover. Measurements of the rapid expansion of the dust front provide a velocity of 25 ms$^{-1}$ ± 2 ms$^{-1}$. The storm grew rapidly zonally and meridionally with dust propagating with velocities V = 19 to 24 ms$^{-1}$ (± 2 ms$^{-1}$) (January 2 and 4, Fig. 8b-c). On January 6 ($L_s = 153°$, sol 313) the dust storm had grown to a band, south of Jezero, with length of 4150 km and width 2000 km, plus an additional north-south branch developing west of Jezero that reached 26°N latitude, surrounding Syrtis Major (Fig. 8d). The storm reached the regional scale at that time, with an area $\sim 8.7 \times 10^6$ km$^2$. At this time, MARCI images showed that the dust was mixed with water ice clouds. The dust began to cover the skies of Perseverance on January 6 (sol 313, $L_s = 153°$, Figure 1d)(Lemmon et al., 2022). This storm is shown in Figure 1 of Wolfe et al. (2023) and in Figure 1c in Guha et al. (2024).

It is in sols 313-315, when the dust optical depth was increasing at Perseverance from $\sim$ 0.3 to 0.8, (Fig 1d) when the amplitude of the diurnal and semidiurnal tides reached their highest values (Fig. 2a-b). The pressure increase was 17 Pa for $S_1$ and 11 Pa for $S_2$. In these sols, the phase of the diurnal tide oscillated between 2 and 3 hr (1/12$^{th}$ and 1/8$^{th}$ of the daily cycle), and that of the semidiurnal tide, between 4-6 hr (1/3$^{th}$ and 1 half of the 12 hour semidiurnal cycle). More detail can be found in Fig. 15 of paper 1.

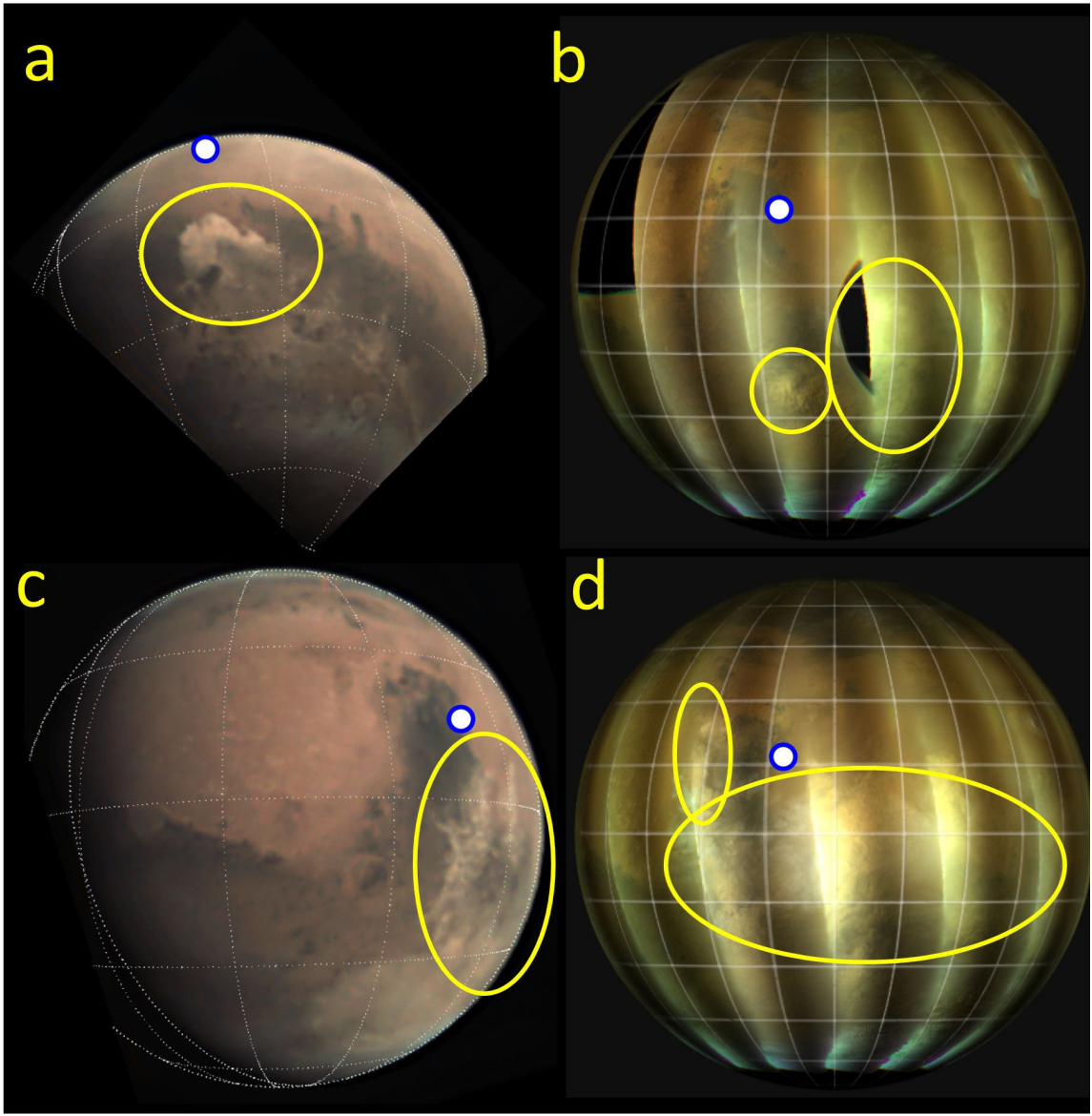

**Figure 8.** *Dust Storm close to Jezero from 1 to 6 January 2022 identified by the yellow oval. (a) 1 January (22:38:25 UT) ($L_s$ = 151.2°, sol 309) VMC/MEX. (b) 3 January (07:41:58 UT) ($L_s$ = 151.8°, sol 310) MARCI/MRO. (c) 4 January (04:38:20 UT) ($L_s$ = 152.3°, sol 311) VMC/MEX. The north-south line on the left marks the terminator. (d) 6 January (03:01:39 UT) ($L_s$= 153.3°, sol 313) MARCI/MRO. Perseverance location is marked by a blue-outlined white disk.*

The dust storm reactivated west of Jezero on 8 January 2022 ($L_s$=154.4°, sol 315) at a location (25°N, 63°E), spanning an area of $4.1 \times 10^6$ km². This dust was transported by westerly winds to cover the skies of Perseverance on sols 315-317 (8-11 January) (Fig. 9). The optical depth reached its maximum value of ~ 1.4 from 9 to 11 January ($L_s$ ~ 155°-156.1°, sols 316-317) (Lemmon et al., 2022, Fig. 1d). No orographic or albedo features are visible in the images of Jezero on these sols (Fig. 9a-c). By 11 January 2022 (sol 318.5, $L_s$ ~ 156.1°) the skies had cleared, the dust had settled, and the surface details were again visible (Fig. 9d).

It is interesting to note that similar dust storms (optical depths > 1 at a wavelength 880 nm) were also observed in different years for this epoch ($L_s \sim 135° – 170°$) by the Mars Exploration Rovers Spirit (MY 27, 28, 29) and Opportunity (MY 27) (Lemmon et al., 2015). The effects of the dust storm in MY 36 produced an increase in the optical depth $\tau \sim 2$ at Gale crater as observed by the Mars Science Laboratory (MSL) (Lemmon et al., 2024).

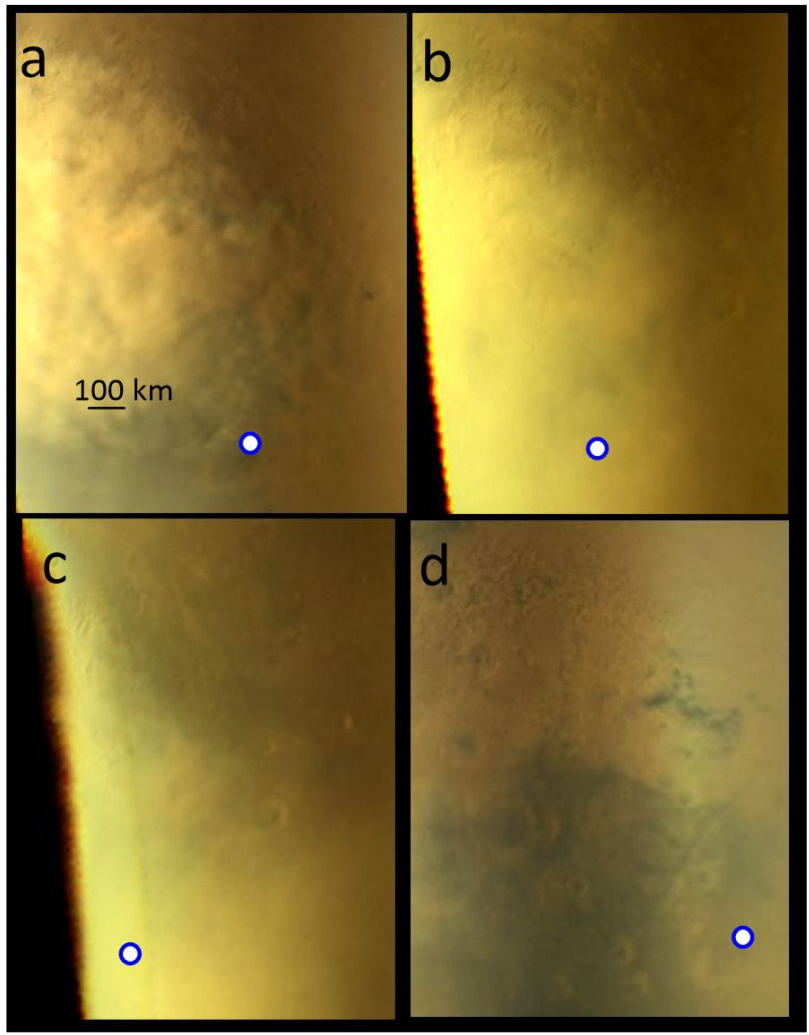

**Figure 9.** *Dust Storm over Jezero from 8 to 11 January 2022. (a) 8 January (11:08:03 UT) (Ls 154.4°, sol 315, LTST 15.9 hr) MARCI/MRO. (b) 9 January (11:26:49 UT) (Ls 155.03°, sol 316, LTST 15.2 hr) MARCI/MRO. (c) 10 January (11:45:38 UT) (Ls 155.56°, sol 317, LTST 15.24 hr) MARCI/MRO. (d) 11 January (13:56:33 UT) (Ls 156.13°, sol 318, LTST 15.2 hr) MARCI/MRO. The Perseverance location is marked by a blue-outlined white disk.*

### *6.5. North Pole edge dust storms ($L_s$ = 160°-190°)*

In the $L_s \sim 160°$-$190°$ period (sols 326 - 379), from 19 January to 14 March 2022, large irregular and elongated dust storms developed at the edge of the North Pole (60°N to 80°N), most of them concentrated in the longitude sector $\sim 330°E - 360°E$ (Fig. 10). These storms had lifetimes between 1-3 sols, moved eastward with translation velocities 6 to 18 ms$^{-1}$ (± 1 ms$^{-1}$) and the most elongated reached lengths of about 4000 km. Their centres were at a distance of 1500 km from Perseverance; while the storm's closest edge was at a distance of 925 km.

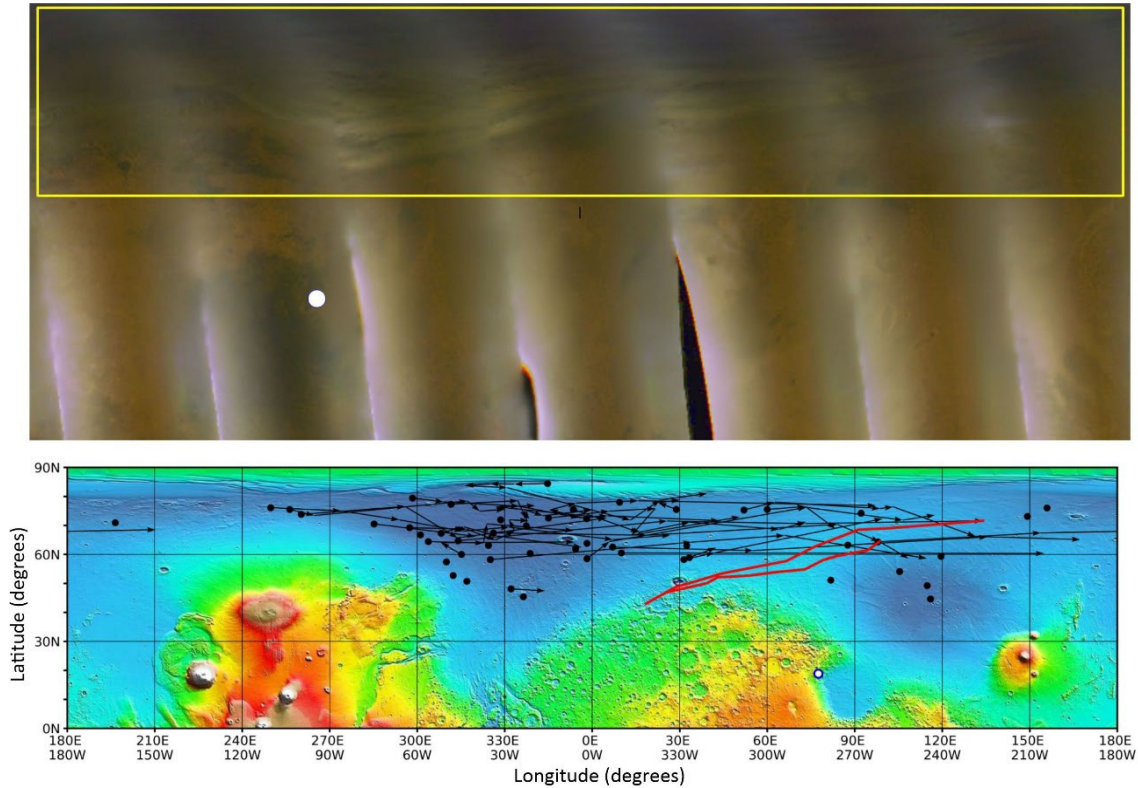

**Figure 10**. *(a) Map from MARCI images showing a large dust storm on 14 March 2022 ($L_s$ = 190.1, sol 379) at the North Pole edge (Perseverance indicated by a white dot). (b) Location (black dots) and tracks (black lines) of the dust storms observed from $L_s$ = 160° to 190° (sol 326-379) at LTST ~ 15.5 hr. The red lines corresponds to two of the most significant cases of elongated storms observed at $L_s$ ~ 180° due to their extension and edge proximity to the rover.*

### 6.6. Regional Dust Storm A ($L_s$ = 200° – 235°)

A yearly recurrent regional dust storm event named A typically starts in the southern hemisphere at $L_s$ ~ 205° (Kass et al., 2016; Guha et al., 2024; Martin-Rubio et al., 2024). This storm propagates northward and produced increases in the optical depth at visible wavelengths that were detected by the Spirit and Opportunity rovers (Lemmon et al., 2015) and by Curiosity (Chen-Chen et al., 2019; Lemmon et al., 2024).

Storm A was detected at Jezero when the optical depth raised above the average value measured in the previous sols, i.e. when $\tau \geq 0.6$, on sol ~ 395 ($L_s$ = 200°, 31 March 2022) (Figure 1d). Optical depths reached values $\tau \geq 1$ between sols 417 and 472 ($L_s$ ~ 213°-249°, 22 April – 18 June 2022). By mid-September 2022 (sol 560, $L_s$ ~ 305°) the surface albedo features recovered their contrast when $\tau$ decreased to ~ 0.5 (Fig. 11).

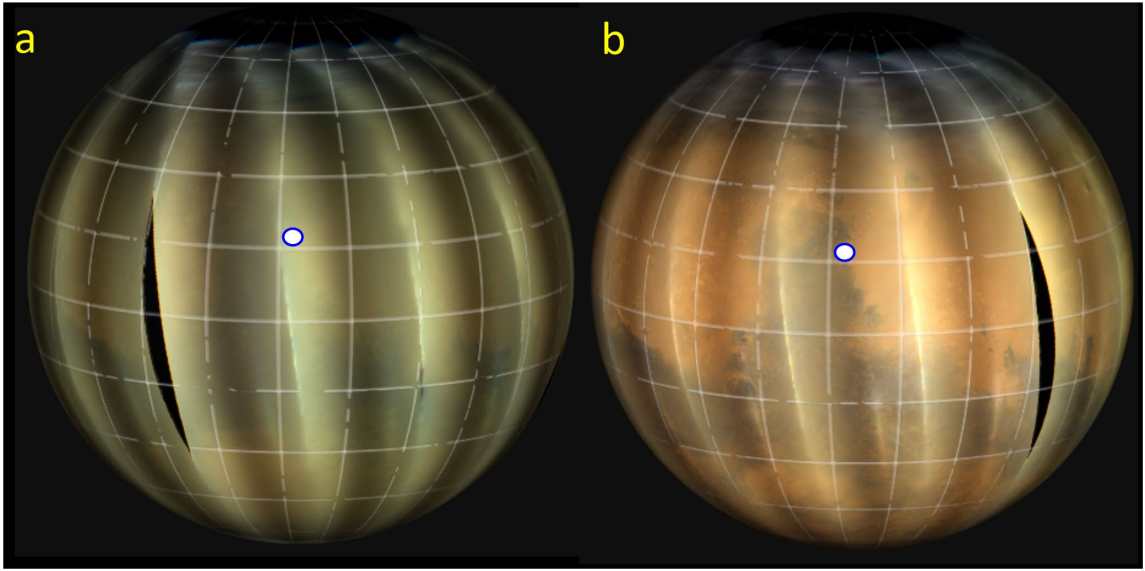

**Figure 11.** *Dust increase (left) and decrease (right) from May to September 2022 during storm A from MARCI/MRO images. (a) 16 May (20:46:17 UT) (Ls 228.6°, sol 440, LTST 15.2 hr). (b) 13 September (05:50:22 UT) (Ls 303.1°, sol 557, LTST 14.06 hr). The blue-outlined white disk marks the Perseverance location.*

The arrival of storm DS-A at Jezero produced large changes in the amplitudes and phases of the tides (Fig. 2a-b, g). The amplitude $S_1$ had a peak within a large fluctuation of ~ 15 Pa (peak-to-peak) between sols 420 and 435 ($L_s$ ~ 215°-225°). Similarly, the amplitude $S_2$ had a peak following a fluctuation of ~ 12 Pa between sols 424 and 428 ($L_s$ ~ 219°). The phase of $S_1$ dropped ~ 4 hr (~ 1/6$^{th}$ diurnal cycle) and that of $S_2$ increased by 3hr (~ 1/4$^{th}$ of the semidiurnal cycle) near simultaneously in sol 427 ($L_s$ ~ 220°). This behaviour repeats at the passage of the storms DS-1, DS-A and DS-C in Jezero, i. e. there is a sharp increase in diurnal and semidiurnal amplitudes (Figures 2a and 2b) and a fall in the diurnal phase with simultaneous increase in semidiurnal phase (Figure 2g) in agreement with what was reported at the Viking location (Leovy and Zurek, 1979). The case of storm DS-NPC is less obvious as discussed in section 6.8.

### 6.7. Regional Dust storm C ($L_s$ = 308°- 318°)

Dust storm C belongs to another seasonally recurring class of regional storms that happen during years without a Global Dust Storm (Kass et al., 2016; Battalio & Wang 2021; Guha et al., 2024; Martin-Rubio et al., 2024). The onset of this storm in MY 36 took place on 21 September 2022 ($L_s$=308°, sol 565) at 14°N and 313°E (North of Vallis Marineris) (Fig. 12). The storm showed a complex behaviour. First, it rapidly grew and expanded East, West and South on 22 September, and then from 23 to 24 September, it propagated southward with a velocity V = -16 ms$^{-1}$ (± 1-2 ms$^{-1}$ for all velocity measurements) reaching the latitude 60°S (Fig. 12). Then, an elongated dust branch moved westward at a latitude of 5°S into Valles Marineris with a velocity of -7ms$^{-1}$ and, simultaneously, expanded along a meridional circle, northward with V = +16 ms$^{-1}$ and southward with V = -26 ms$^{-1}$. From the 24 to the 29 of September, a dust tongue south of the Equator (12°S to 25°S), propagated eastward with a velocity of 20 ms$^{-1}$. Finally, another dust branch expanded westward, parallel to the south polar cap along latitudes 50°S to 60°S, with a velocity -8 ms$^{-1}$. By 29 September (L$_s$=312.5°, sol 572) the dust storm covered about 1/5$^{th}$ of the Martian surface (area ~ 2.9x10$^7$ km$^2$), extending from latitudes ~ 30°N to ~ 62°S and from longitudes 240°E to 99°E (Fig. 12).

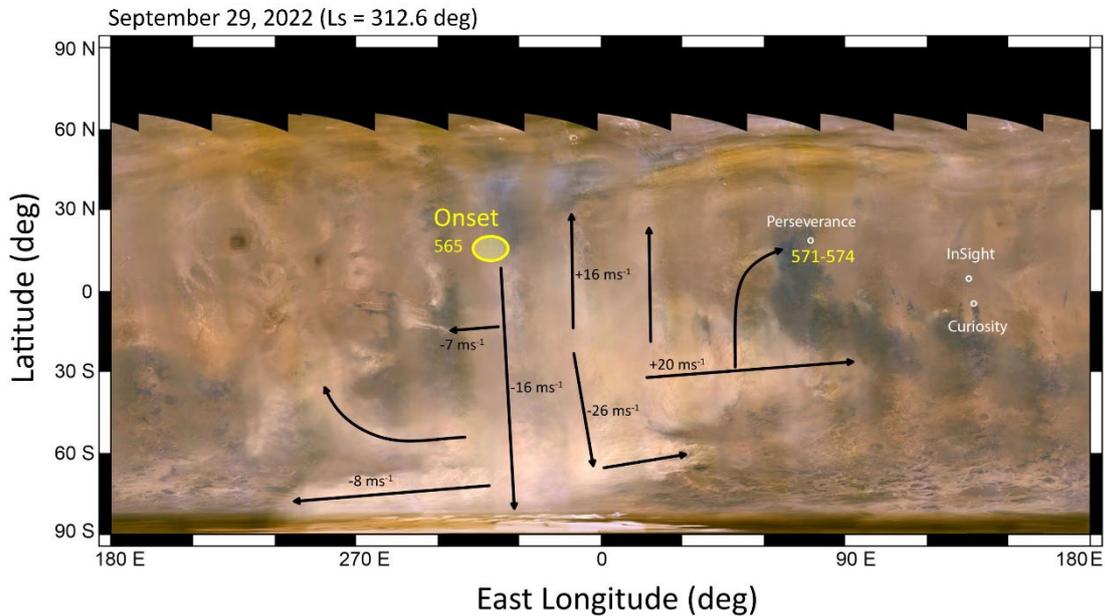

**Figure 12.** *Map from 29 September 2022 (sol 573, L$_s$=312°), showing the dust coverage during the advanced phase of regional Dust Storm C, 8 days after its onset (L$_s$=308°, sol 565, indicated by the yellow oval). The arrows mark the main directions of dust propagation with the measured velocities (+ for northward and eastward motions, - for southward and westward motions). Typically the velocity uncertainty is ± 1 ms$^{-1}$. The sol number is indicated in yellow for the onset and dust arrival at Perseverance. The map comes from MARCI/MRO, NASA PIA 25412.*
*([https://photojournal.jpl.nasa.gov/catalog/PIA25412](https://photojournal.jpl.nasa.gov/catalog/PIA25412)).*

VMC images obtained at these dates show the dust covering Perseverance (Fig. 13a). The sequences of overexposed images taken near MEX orbital pericenter allow us to observe the brightness of the dust in twilight and over the limb (Fig. 13b). Using the method described in Hernández-Bernal et al. (2019), we measured the height of the dust storm tops. We found heights of ~ 30 km in the twilight region with an inhomogeneous distribution of dust along the south to north direction (points 2-4 in Fig. 13b). At the limb (point 1 in Fig. 13b), the dust formed a detached layer over the western flank of the Hellas basin, where it reached higher altitudes of ~ 60-80 km. These data show the horizontal and vertical inhomogeneity in the dust distribution.

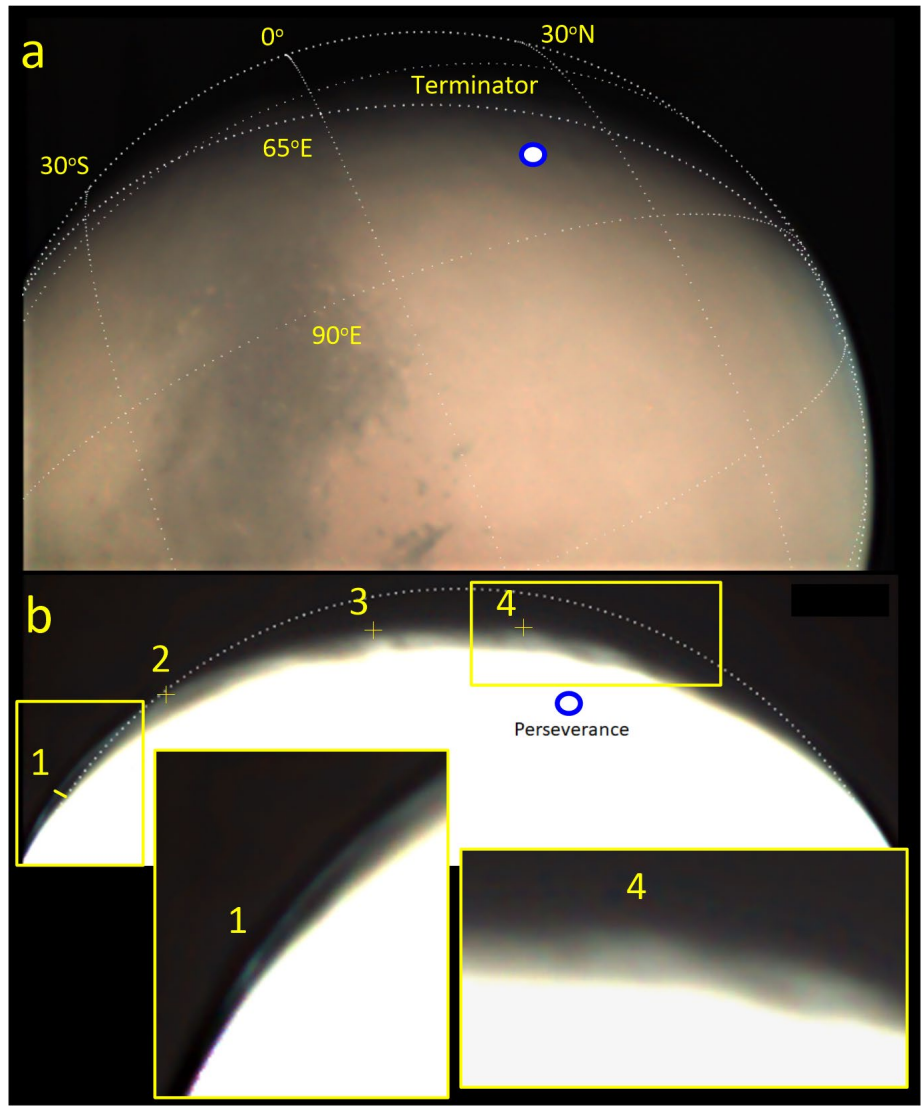

**Figure 13.** *Regional dust storm C imaged on 4 October 2022 (sol 577.5, $L_s$ = 315.5°) by VMC/MEX. (a) Image obtained at 8:57 UT with the dust extending north of the equator covering Perseverance location (marked by a blue-outlined white disk). (b) Overexposed image showing the dust above limb (point 1) and in the twilight beyond the terminator line (points 2, 3, 4). The insets show an augmented view of the area around points 1 (above limb) and 4 (beyond terminator). The location of these points is, 1 (50°E, 40°S), 2 (42°E, 20°S), 3 (57°E, 4°S), 4 (62°E, 20°N). In (a) and (b) the external dashed circle marks the planet's limb.*

The optical depth started to increase in Perseverance on sol 571 (28 September 2022, $L_s$ 312°), i.e. six sols after the DS-C onset, and reached a maximum optical depth on sol 574 (1 October) (Fig. 1d and 14). The presence of dust produced large and complex changes in the daily pressure cycle that strongly manifested in the amplitude and phase of the diurnal and semidiurnal tides (Figs. 2a-b, 2g, 14a-b). For example, the amplitude of the diurnal tide $S_1$ started to increase in sol 567 from a previous average value of 10 Pa to a maximum of 23 Pa in sols 569-571 (Fig. 14a). Simultaneously, it suffered a full change in the phase (24 hr or 360°) in sols 571-574 (Fig. 14b). These changes in $S_1$ occurred 3-4 sols before the optical depth reached its maximum value at Perseverance in sol 574, i.e. while the dust was still propagating from the west toward Jezero (section 6.7, Fig. 12). A similar behaviour was reported at Gale crater (Fig. 11 in Zurita-Zurita et al., 2022) suggesting that this is a robust dynamical coupling on a planetary scale. The semidiurnal amplitude underwent a drop of ~ 2 Pa in $S_2$ in sol 568, i.e. 3 sols after the onset of the storm, when the dust was still far from Jezero. This was followed by a sharp increase in $S_2$ to about 21 Pa, reaching the maximum in sol 576, while its phase shifted by 2-3 hr between sols 567 and 579. The changes in the phases of the diurnal and semidiurnal components occurred simultaneously (Fig. 14b), but the maximum in $S_1$ occurred about 8 sols earlier than that of the $S_2$ (Fig. 14a). The storm also affected the amplitude and phase of the terdiurnal component $S_3$, but with small changes, close to the detection limit of the retrieval.

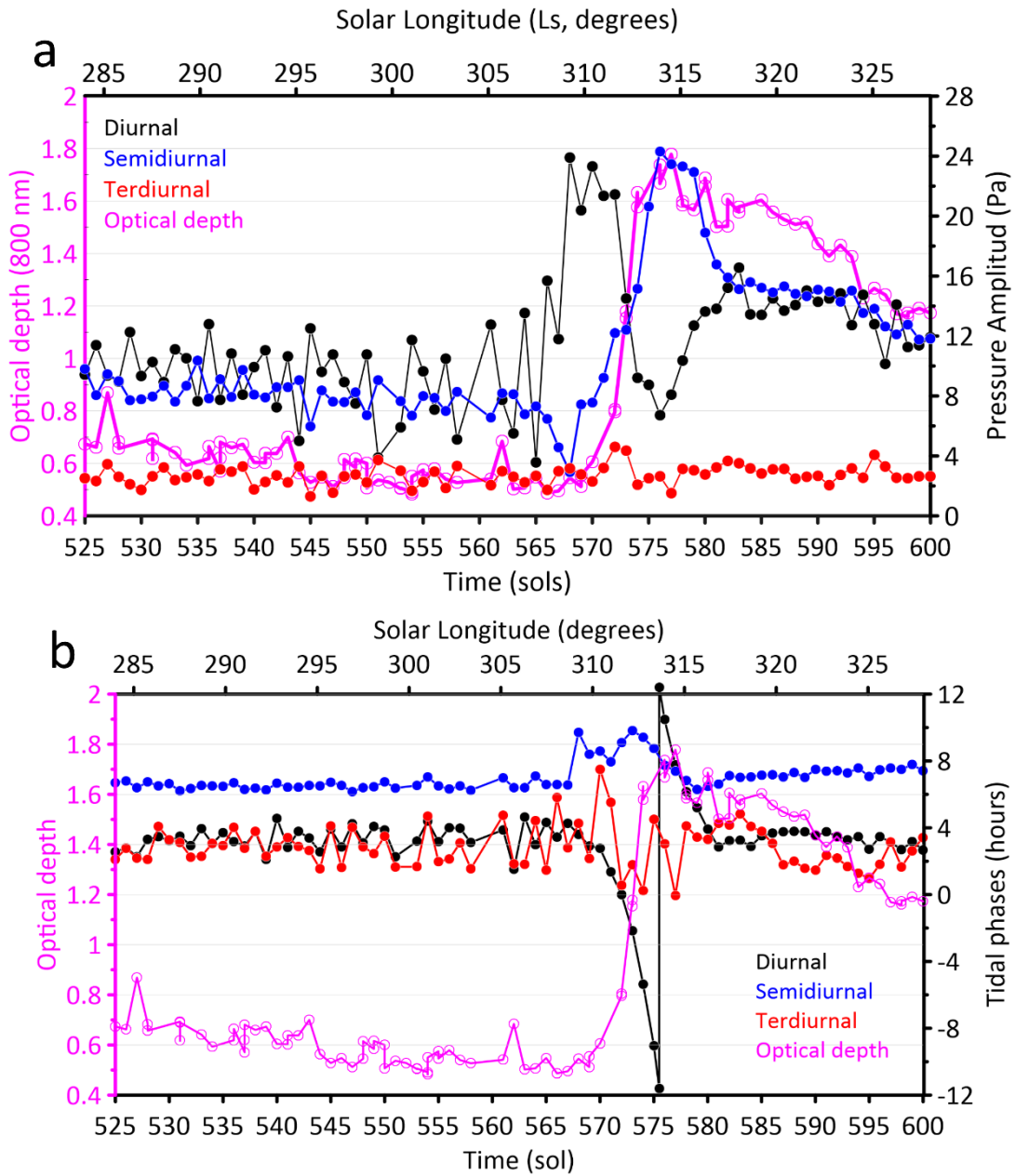

**Figure 14**. *Changes in the amplitude and phase of the tides during the evolution of Dust Storm C. (a) Amplitude of the diurnal, semidiurnal and terdiurnal components together with the optical depth evolution. (b) Same as (a) but for the tidal phases.*

### 6.8. North Polar edge dust cyclones ($L_s$ =335°-360° and 0°-17°)

This is the start of a new season of formation of dust cyclones at the edge of the NPC, an activity that continues up to $L_s$ = 90° (see section 6.1). The first storm of the season was observed on 11 November 2022 (sol 615, $L_s$ = 336.8°) centred at about 52°N and 162°E. Subsequently, new cyclonic storms were regularly observed at the edge of the NPC (see another example in Gebhardt et al., 2023). The most interesting cases are those in which the storm comes close enough to Jezero to introduce dust into the skies over Perseverance or to disturb the pressure measurements. One such case occurred between 18 and 27 December 2022 (sols 650 – 659, $L_s$ = 356° - 0.7°) (Figure 15). A storm that we called DS-NPC grew up north of Perseverance, evolving from a compact textured feature to an arc-shape storm. The disturbance moved eastward at 15 ms$^{-1}$ and its centre was ∼ 1480 km north of Perseverance on sols 654-655 (Fig. 15 (b) and (c)). The edge of this storm was only 585 km from Perseverance, close enough to inject dust over the rover. On sols 657-658 the optical depth data show a peak in opacity (Fig. 1d) and the diurnal and semidiurnal amplitudes of the thermal tides raised their pressure by ∼ 9 Pa (sol 659, $L_s$ =0.4° in MY37) and by ∼ 2 Pa (sol 655, $L_s$ =358.4° in MY36) (Fig 2a-b), respectively. Although smaller in amplitude, this behaviour remembers what was seen with major storms. However, their phases do not show a change in sols 650-660. There is a previous peak in opacity in sols 600-640, with a corresponding increase in the amplitude of the diurnal component, this time leaving no clear signal in the semidiurnal tide or in the phases. We did detect a tidal phase trend similar to previous storms in sol 703 ($L_s$ ∼ 22°), with the diurnal phase dropping ∼ 2 hr (∼ 1/12$^{th}$ diurnal cycle) and semidiurnal increasing by 3hr (∼ 1/8$^{th}$ diurnal cycle) simultaneously (Figure 2a, 2b and 2g). We think these phase changes are not related to storm DS-NPC because the time difference is around 40 sols. Later on, other storms formed regularly at NPC and were observed up to the end of our imaging survey on 31 January 2023 ($L_s$ = 17.4° in MY37, sol 693). It should be noted that these dust storms are also accompanied by clouds, probably formed along with the storm but at higher altitudes than where the dust is injected (see section 6.1, Figure 4).

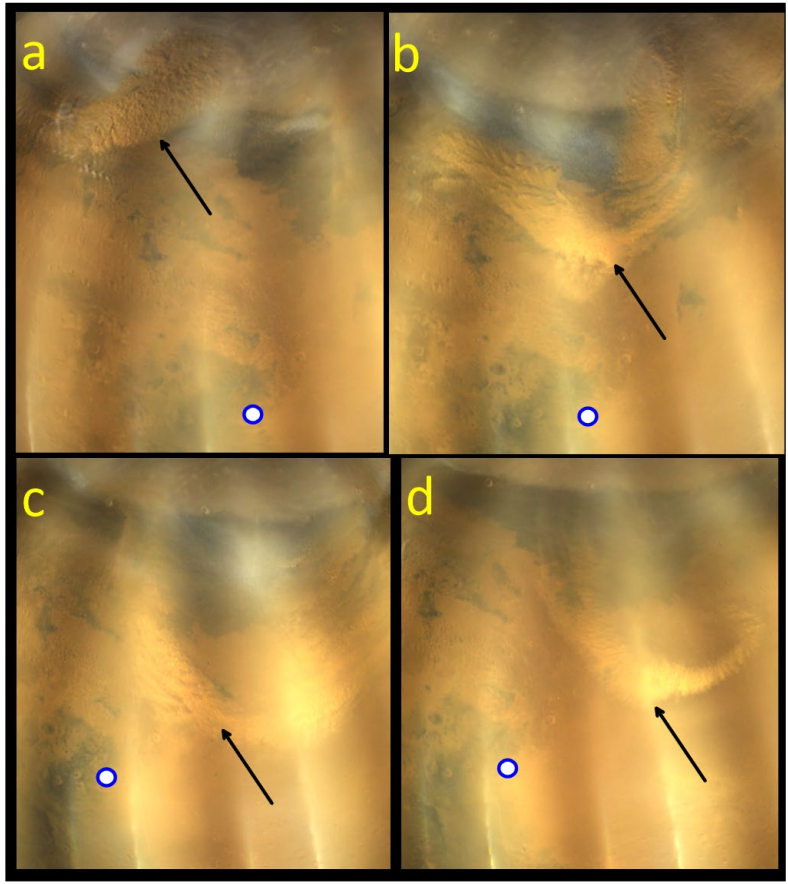

**Figure 15.** *A compact dust storm (arrow) evolving to an arc-shape feature approaching Perseverance on MARCI images. (a) 21 December 2022 (Ls= 357.7, sol 653); (b) 22 December 2022 (Ls= 358.2, sol 654); (c) 23 December 2022 (Ls= 358.7, sol 655); (d) 27 December 2022 (Ls= 360.7, sol 659). In all these images LTST ~14 hr. The blue-outlined white disk marks the location of Perseverance.*

**7. Discussion**

In this section, we study the effects and correlations between the synoptic and planetary-scale disturbances presented in section 6 and the long-period waves and tides as characterized from the surface pressure measurements by Perseverance presented in sections 3-5.

***7.1 Dust and cloud optical depth and their effects on tides***

We first examine the relationship between the amplitudes of the thermal tides and the aerosol content in the atmosphere as measured at Jezero. Figure 16a shows the comparison between the normalized amplitudes of the diurnal and semidiurnal components (i. e. the tide amplitude divided by the mean pressure in that sol) with the aerosol optical depth according to the empirical formulation proposed by Wilson et al. (2008) (and paper 1). The correlation between the seasonal variation of $S_1$ and $S_2$ with a linear function on τ is good for $L_s$ ~ 130° - 360°, i.e. during the dusty period. However, no correlation is found in the range $L_s$ ~ 0° - 130° when $S_1$ and $S_2$ show opposed seasonal trends during the low-dust period, when the contribution to the optical depth of the water-ice clouds was significant (Toledo et al., 2023; Smith et al. 2023; Patel et al., 2023), and the total τ ~ constant (Fig. 1d).

The relationship between the seasonal variation of the combined tidal amplitudes and the total optical depth measured at Jezero is further explored in Fig. 16b. Here, as in paper 1, we show that the observed pattern in the seasonal evolution of the optical depth is similar to that followed by a combination of a 50% of $S_1$ and $S_2$ components. This correlation slightly improves when the combination includes a mixture of $S_1$ to $S_4$ amplitudes in the following percentages: $S_1$ (27%), $S_2$ (53%), $S_3$ (16%) and $S_4$ (4%).

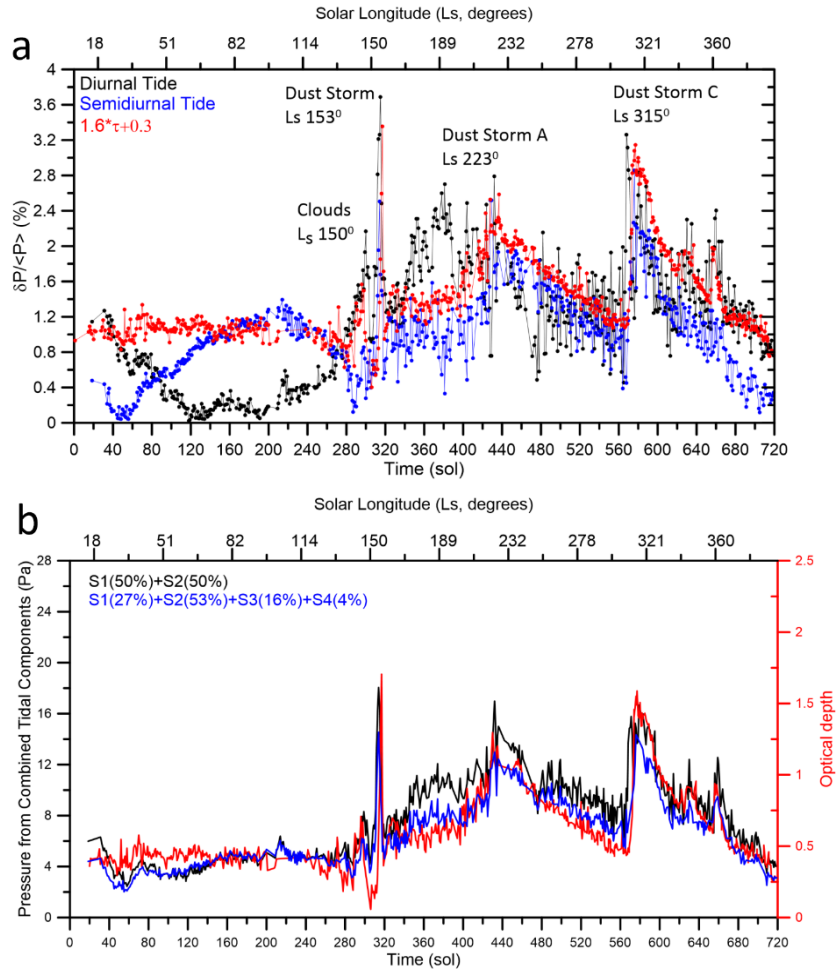

**Figure 16.** *Relationships between the amplitudes of the thermal tides and the aerosol optical depth. (a) Normalized amplitudes of the diurnal $S_1$ and semidiurnal $S_2$ and their relation with a linear function of the optical depth given by $1.6\ \tau + 0.3$. (b) Relationship between the optical depth and two combinations of the amplitudes of the tides.*

### *7.2 Long-period waves*

We have seen in sections 6.1, 6.2 and 6.8 that abundant cyclones develop at the NPC edge during the $L_s$ periods 335°-360° and ~0°-140° (approximately the epoch of the ACB development). They evolve in the latitude band from ~ 55°N to 75°N and follow the polar ice-cap retreat (Figure 5). Most of these cyclones generate strong surface winds that lift the dust up to 6-11 km (Sánchez-Lavega et al., 2022). Other cyclones, as the recurrent DAC ($L_s \sim 120°$-$140°$), do not lift dust but form clouds at heights ~ 10-20 km (Sánchez-Lavega et al., 2018b). If the observed pressure oscillations at Jezero, with amplitudes between 2-4 Pa and periods between 3-4 sols (Figure 3), are related to this cyclone activity, the radius of action of the low pressure disturbance associated with the cyclonic vorticity would be larger (by a factor 2-3) than the vortex size as traced by dust and clouds. From the measured size of the observed cyclones L ~ 1,000 – 2,500 km we get wavenumbers $n \sim (\pi R_M \cos 60^0)/L \sim$ 2-5 (assuming the cyclone size is half a wavelength) where $R_M$ is the radius of Mars. If we use the radius of action of the depression (indicated above) instead, then $n \sim$ 1-3 which is in agreement with previous works (Barnes et al., 2017; Collins et al., 1996; Hinson and Wilson, 2021). Waves with periods in the range 2-5 sols (and wavenumbers 1-3) were also inferred from the signal lag in the pressure data measured at the same season at Jezero and at Gale crater by Curiosity (Battalio et al., 2022).

The amplitude of the pressure oscillations increased dramatically around sol ~ 300 ($L_s \sim$ 146°) with the start of the dusty season, and particularly during the evolution of dust storms over Jezero, when peak to peak amplitude were above 10 Pa (Figure 3). The largest amplitude was ~ 22 Pa, and it was measured during the dust storm C on sols ~ 560 to 580 ($L_s$ = 307°-312°) just before the dust reached its maximum optical depth at Jezero (Figures 1d, 14a).

Battalio and Wang (2020) studied the evolution of eddy activity during large-scale dust storms. Here, we have analyzed the behavior of the pressure oscillations along the Martian Year 36 at Perseverance location by conducting simulations with the LMD Martian PCM (Planetary Climate Model)(Forget et al., 1999) following the method described in Lewis et al. (1999) and Millour et al. (2022). Figure 17 shows pressure oscillations as obtained from the simulation in the region of Jezero crater when filtering the pressure data from the diurnal mean and seasonal trend. The simulation described in this section were carried out with a 64x48 lat/lon grid (2.8°x7.5°, latitude x longitude), and 36 vertical levels, with a vertical resolution decreasing from 5 m near the surface to 10 km at the model top (~80 km). The physical time-step is ~7 min and we use windows of 10° in $L_s$ (in steps of 0.5°). The reference simulation is performed with two-moment aerosol sizes distributions for water ice and dust. Dust particles are represented by a lognormal particle size distribution with an effective variance of 0.5 and an effective particle radius of 3 μm. This choice is driven by better agreement between simulated and observed opacities. The simulation has been carried out over multiple annual cycles for MY36 using the available MCS- and EMIRS-derived column opacity maps for that year (Montabone et al., 2023), as a constraint for the simulation to match. In practice, dust is injected from the surface into the PBL when the simulated dust column opacity is lower than that in the column opacity map so that the aerosol and temperature distributions could reach a seasonally equilibrated state. We use radiatively active dust, water vapor and water ice clouds. The period of the model simulated oscillations range from ~ 2 sols (sol range ~ 380 – 470, Ls=190°-248°) to ~ 8 – 20 sols (sol range 180 – 290, Ls=89°-141°).

Overall, the model reproduces the main trend observed with MEDA (Figure 3), with low amplitude oscillations obtained during the first half ("clear season") of the Martian year and larger amplitude oscillations obtained during the second half ("dusty season"). The seasonal trend in the peak-to-peak periods and amplitudes predicted by the model agrees well with that observed for the range of sols ∼ 10-150 (with periods 4 - 5 ± 1.5sols) and ∼ 340-700 (with periods 3 - 4 ± 1 sols). As observed, the atmospheric waves have large amplitudes during the dusty season. It could be tempting to assign the oscillations observed during the period from $L_s$ ∼ 337° (sol 615) to $L_s$ ∼ 44° (sol 80, MY36), to the transient dust cyclone activity at the NPC edge (Figs. 4, 5, 14). The model and observations are consistent in this period within the global trend, although the model predicts higher amplitudes than observed (Fig. 3 and 17). Clarification of this point would require new simulations, which is beyond the scope of this paper.

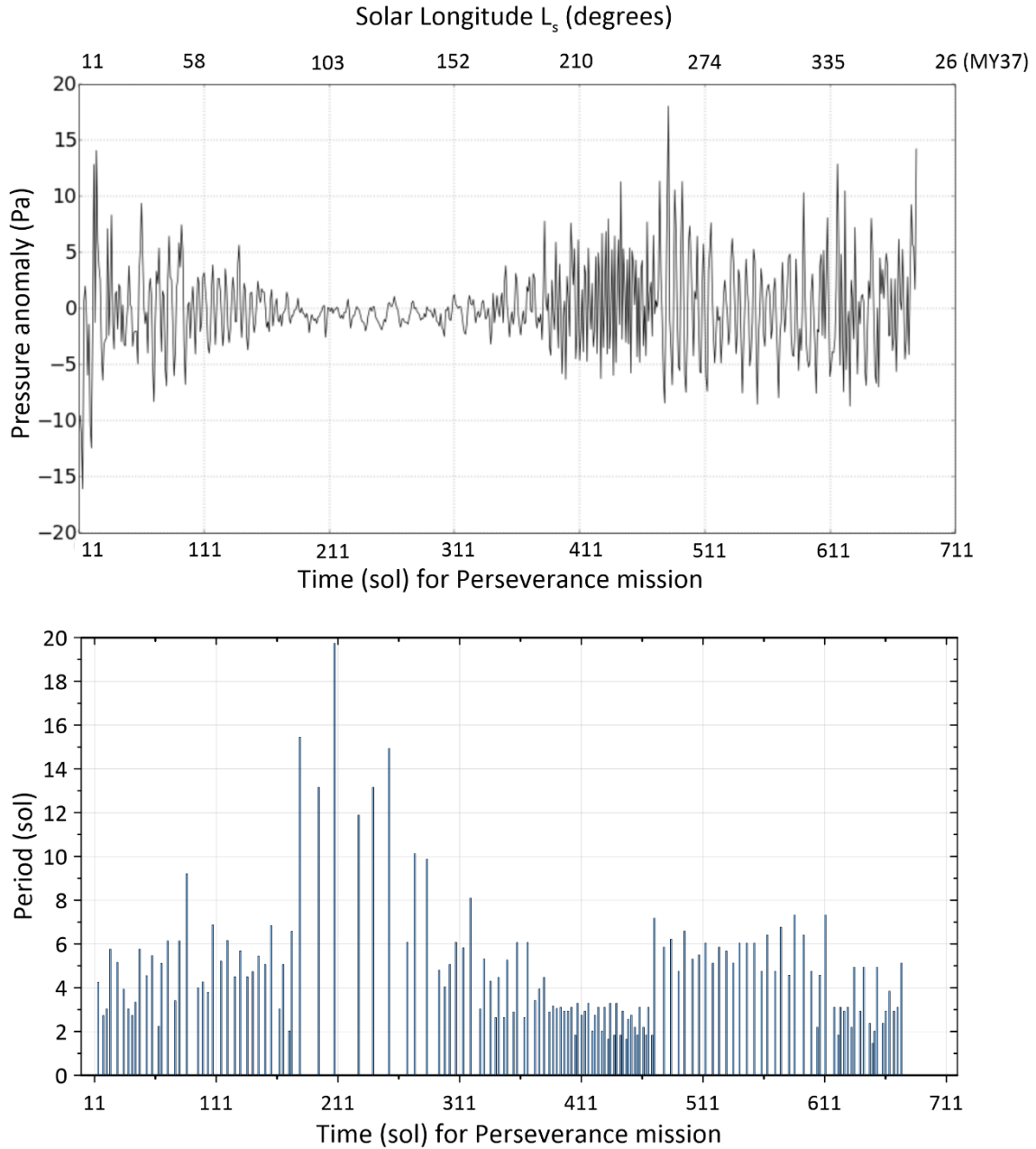

**Figure 17.** Pressure oscillations obtained from a PCM simulation in MY36 in the region of Jezero crater, calculated as residuals between the simulated mean daily pressure and the mean pressure over a 10-sols sliding window (seasonal trend). Upper plot shows the amplitude and the lower plot the period. The resulting oscillations are mostly due to the baroclinic activity.

Baroclinic instability induces waves with periods < 8 sols as observed by sensors on landers (Banfield et al., 2020; Barnes, 1980, 1981, 1984; Barnes et al., 1993; Haberle et al., 2018; Zurita-Zurita et al., 2022). The observations and measured properties of cyclones from orbit (Banfield et al., 2004; Battalio, 2022; Greybush et al., 2019) are compatible with their baroclinic origin (Hunt & James, 1979; Barnes, 1984; Sánchez-Lavega et al., 2018b, 2022; Battalio & Wang, 2020; Hinson & Wilson, 2021). Here, we use the maximum growth rate of baroclinic disturbances in Mars' northern hemisphere to assess the development of the observed cyclones at the North Polar Cap edge. A useful estimate of the growth rate is given by the baroclinic index, defined as (Lindzen & Farrell, 1980; James & Grey, 1986; Lembo et al., 2017; Battalio et al., 2016):

$$\sigma_{BI} = 0.31 \frac{f}{N} \left| \frac{\partial U}{\partial z} \right| \qquad (1)$$

Here $f = 2\Omega \sin\varphi$ is the Coriolis parameter with $\varphi$ the latitude and $\Omega$ =7.08x10$^{-5}$ s$^{-1}$ the angular rotation velocity of Mars, $N^2(z) = \frac{g}{T(z)} \left[ \frac{dT}{dz}(z) + \frac{g}{C_p} \right]$ is the Brunt-Väisälä frequency where we used for the adiabatic gradient $g/C_p$= 4.5 K km$^{-1}$, and $\frac{\partial U}{\partial z}$ is the vertical shear of the zonal wind velocity (*U*). Although equation (1) was derived for a constant vertical wind shear, Lindzen and Farrell (1980) argue its more general validity. We preserve in (1) the coefficient 0.31 to get the approximate time-scales involved in the growth rate of the baroclinic disturbances in Mars. To calculate the baroclinic index $\sigma_{BI}$, we use the MCD for standard climatology and average solar conditions to get the temperature and zonal wind maps (latitude-altitude), *T*(*z*,$\varphi$) and *U*(*z*,$\varphi$), at the Perseverance longitude and for LTST ranging from 7 hr to 12 hr. We selected the values of $L_s$ for representative cases of the observed wave activity. We then calculate the mean *N*(*z*) and (*dU/dz*) (*z*) in the latitude range ∼ 55°N - 65°N where the peak velocity of the northern eastward jet stream is predicted.

Figure 18 shows the vertical profiles of the growing time of the instability (the inverse of the baroclinic index $\sigma_{BI}$) for the selected $L_s$ periods at 10:00 LTST. Globally, we can divide the behavior of the growing time profiles in two periods. In the first part of the year, for $L_s$ ∼ 45° to 135°, the fastest growing time occurs at two heights ∼ 4 km and 22 km. Closer to the surface, the temperature profile in the model is convectively unstable even at 07:00 LTST in the morning. The observation of the dust spirals and similar features in this epoch (Fig. 4-5 and 15), suggests that the baroclinic instability actually extends near the surface where the dust cyclones form and the vortex winds raise the dust (Hollingsworth et al., 2014; Mulholland et al., 2016). During the dusty period ($L_s$ ∼ 180°-360°), the amplitude of the surface pressure oscillations measured by Perseverance increases as does the amount of dust in the atmosphere (Figs. 3, 17). The growing time of baroclinic waves in this period has its fastest values very close to the surface at LTST 7-12 hr and is quite constant with altitude above 4 km. This time of the year corresponds to the polar night in the northern hemisphere and to the development of the North Polar Hood..

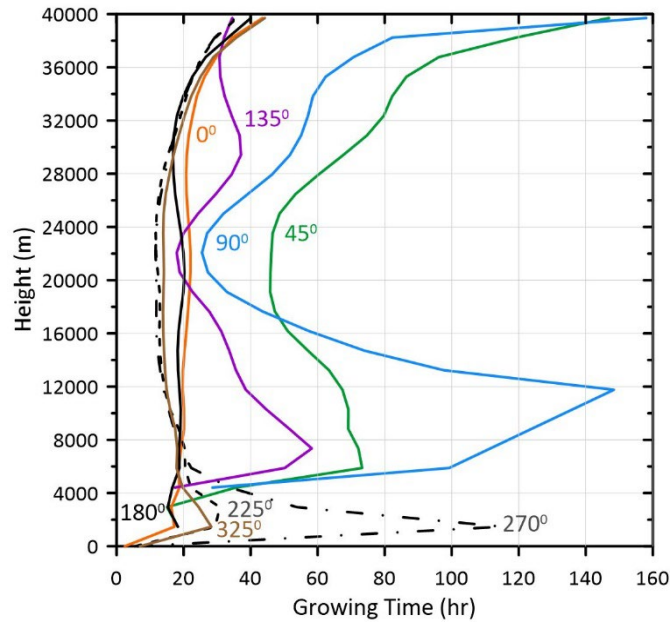

**Figure 18**. *Vertical profiles of the growing times of baroclinic disturbances associated to the northern eastward jet stream centered at latitude ~ 60°N at the longitude of the Perseverance rover (77.5°E) and at LTST = 10 hr for selected values of the solar longitude $L_s$ (based on MCD data).*

## 8. Summary

We have used the Martian year 36 as a case study of the relationship between surface pressure measurements and the development of medium and large scale atmospheric phenomena in the northern hemisphere of Mars observed in orbital images. Our most relevant findings are:

- Thermal tides: In the dusty period (broadly, $L_s$ ~ 130°-360°), the amplitudes of the diurnal and semidiurnal components of the daily pressure cycle correlate with the optical depth measured at Jezero and with the development of dust storm events near Jezero. No correlation is found from $L_s$ ~ 0° to 130°, coinciding with the development of the ACB. The amplitude of the tidal components 3 and 5 on the one hand, and 4 and 6 on the other, follow a similar behavior. The seasonal evolution of the aerosol optical depth correlates well with a combination of 50% of the amplitudes of the diurnal and semidiurnal components. The correlation improves when combining the amplitudes of the four components with various empirical weights dominated as before by the contribution of the semidiurnal and diurnal tides. The increase in optical depth produced by water ice clouds correlates well with an increase in the amplitude of the diurnal component and with a pronounced drop in the semidiurnal component.

- Dust Cyclones. Spiral, irregular and comma-arcs vortices traced by dust were observed at the edge of the North Polar Cap (50°N-80°N) in the $L_s$ ranges ~ 0°-90°, 160°-190°, 335°-360° (MY 36) and 0°-17° (MY 37). Their typical size is L ~

1,000 -2,000 km and the most active area where they grow is the Acidalia Planitia (longitude ~ 330°E). These vortices evolve at distances from Perseverance ~ 1,500 km (closest) to 4,500 km (on average). They are most probably behind the pressure oscillations detected at Perseverance with amplitudes of ~ 1.5-4 Pa (amplitudes increasing with dust content in the atmosphere) and periods 2-4 sols. One of them (DS-NPC at $L_s$ ~ 357°-360°) reached Perseverance, leaving its imprint in the diurnal and semidiurnal tides. Their properties agree with a baroclinic origin and their amplitudes and periods with the Martian PCM predictions.

- Double Annular Cyclone (DAC). The recurrent double-annular cyclone, traced by water ice particles, was observed from $L_s$ ~ 120° to 140° at subpolar latitudes near 60°N. This peculiar disturbance, of baroclinic origin, grows and dissipates within each sol, while it translates slowly eastward, evolving at large distances from Perseverance (4,500 km to 7,000 km along a latitude circle). The pressure oscillations detected by Perseverance in this period present a small amplitude (< 2 Pa) and it is uncertain whether the DAC contributes to them.

- Aphelion Cloud Belt. The development of the ACB and the cloud arrival at Jezero, possibly affected the diurnal and semidiurnal tides. The diurnal component was null at $L_s$ ~ 60° - 67° and in $L_s$ ~ 93° - 97° but reached a maximum at $L_s$ ~ 146° when clouds were over Jezero, a trend consistent with measurements of the water-ice cloud optical depth. The phase of the diurnal component underwent a significant decrease of about 8 hr (1/3$^{rd}$ of the daily cycle) from $L_s$ ~ 66°-102° and then a maximum at $L_s$ ~ 114°-139°. The amplitude of the semidiurnal component was zero at the beginning of the ACB at $L_s$ ~ 28°-33° and at its end at $L_s$ ~ 139°-141°, and has a maximum in between. In parallel the phase showed a drop of 4 hr (1/6$^{th}$ of the daily cycle) starting at $L_s$ ~ 112° and with a pronounced minimum at $L_s$ ~ 139° when clouds arrived at Jezero. Comparatively, the behaviour of the diurnal and semidiurnal components, both in amplitude and phase, followed a complementary trend.

- Dust Storms on Jezero. Four episodes injected dust at the location of Perseverance. (a) The first regional storm (DS-1) took place from $L_s$ ~ 150° to 156°. Expansion velocities were of around 20-25 ms$^{-1}$, reaching an area of 4.1-8.7x10$^6$ km$^2$. (b) The recurrent event A (DS-A) took place between $L_s$ ~ 205°-235°. (c) The recurrent storm C (DS-C) took place between $L_s$ ~ 308°-318° showing a complex expansion behaviour with velocities between 16 and 26 ms$^{-1}$ (± 1-2 ms$^{-1}$), reaching a regional scale with an area ~ 2.9x10$^7$ km$^2$. The dust altitude was ~ 30 km at the latitude of Perseverance but in the south it reached 60-80 km. (d) The fourth was a spiral storm (DS-NPC) that evolved at the edge of the North Polar Cap at $L_s$ ~ 1 (MY37).

- Dust Storms and tides. The four dust storms produced first a decrease and then a rapid increase in the amplitude of the diurnal and semidiurnal tides (by a factor ~ 2, or ~ 10-20 Pa). The phase of the diurnal component showed punctual decreases during the storms (~ 2-3 hr, 1/12$^{th}$-1/8$^{th}$ of the daily cycle) except for the case of DS-C that produced a full phase change of 24 hr. The phase of the semidiurnal component showed punctual increases during the four storms of ∼ 3 hr (1/8$^{th}$ of the daily cycle). In some cases, lags of 1-3 sols occur between the maximum in the optical thickness and the maximum change in the tidal amplitudes and phases.

Future work will aim to interpret the phenomena described here and their effect on surface pressure measurements by means of numerical models. Among other phenomena, we need to understand the relationship between the vertical and horizontal distribution of suspended dust and water-ice clouds, the dust storm properties, and the amplitudes and phases in the surface pressure associated with the components of the thermal tides. Another future study is the development of models for the baroclinic cyclones generated at the edge of the NPC (such as the dust spirals and the DAC and other similar vortices) and the structure of their pressure field at large distances.

## Acknowledgements

This work has been supported by Grant PID2019-109467GB-I00 funded by MCIN/AEI/10.13039/501100011033/ and by grant PID2023-149055NB-C31 funded by MICIU/AEI/10.13039/501100011033 and FEDER, UE, and by Grupos Gobierno Vasco IT-1366-19. Parts of this work were also funded by the Aula EspaZio Gela which is supported by a grant from the Diputación Foral de Bizkaia (BFA). EL and JHB were supported by ESA Contract No. 4000118461/16/ES/JD, Scientific Support for Mars Express Visual Monitoring Camera and through the Faculty of the European Space Astronomy Centre (ESAC) - Funding reference ESAC-531. The authors are very grateful to the entire Mars 2020 science operations team.

## Data availability

The images from the VMC/MEX and MARCI/MRO cameras can be retrieved from the references MARCI images (2024) and VMC images (2024). The MEDA/Perseverance pressure measurements are available through Rodriguez-Manfredi, J. A., & de la Torre Juarez, M. (2021). The software is available through the references ISIS-USGS(2024) and QGIS (2024). The Mars Climate Database (MCD) from LMD can be accessed at https://www-mars.lmd.jussieu.fr/mars/access.html.

## References

Banfield, D., Conrath, B. J., Gierasch, P. J., Wilson, R. J., & Smith, M. D. (2004). Traveling waves in the martian atmosphere from MGS TES Nadir data. *Icarus, 170*(2), 365–403. https://doi.org/10.1016/j.icarus.2004.03.015

Banfield, D., Spiga, A., Newman, C., Forget, F., Lemmon, M., Lorenz, R., et al. (2020). The atmosphere of Mars as observed by InSight. *Nature Geoscience*, *13*(3), 190–198. https://doi.org/10.1038/s41561-020-0534-0

Barnes, J. R. (1980). Time spectral analysis of midlatitude disturbances in the Martian atmosphere. *Journal of the Atmospheric Sciences*, *37*(9), 2002–2015. https://doi.org/10.1175/1520-0469(1980)037<2002:tsaomd>2.0.co;2

Barnes, J. R. (1981). Midlatitude disturbances in the Martian atmosphere: A second Mars year. *Journal of the Atmospheric Sciences*, *38*(2), 225–234. https://doi.org/10.1175/1520-0469(1981)038<0225:mditma>2.0.co;2

Barnes, J. R. (1984). Linear baroclinic instability in the Martian atmosphere. *Journal of the Atmospheric Science*s, *4*1(9), 1536–1550. https://doi.org/10.1175/1520-0469(1984)041<1536:lbiitm>2.0.co;2

Barnes, J. R., Pollack, J. B., Haberle, R. M., Leovy, C. B., Zurek, R. W., Lee, H., & Schaeffer, J. (1993). Mars atmospheric dynamics as simulated by the NASA Ames general circulation model 2. Transient baroclinic eddies. *Journal of Geophysical Research, 98*(E2), 3125–3148. https://doi.org/10.1029/92JE02935

Barnes, J. R., Haberle, R. M., Wilson, R. J., Lewis, S. R., Murphy, J. R., & Read, P. L. (2017). The global circulation. In R. M. Haberle, R. Clancy, F. Forget, M. D. Smith, & R. W. Zurek, (Eds.), *The atmosphere and climate of Mars* (pp. 229–294). Cambridge University Press. https://doi.org/10.1017/9781139060172.009

Battalio, M., Szunyogh, I., Lemmon, M.: Energetics of the martian atmosphere using the Mars Analysis Correction Data Assimilation (MACDA) dataset, Icarus, 276, 1-20, 2016. https://doi.org/10.1016/j.icarus.2016.04.028

Battalio, M., & Wang, H. (2020). Eddy evolution during large dust storms. *Icarus, 338*, 113507. https://doi.org/10.1016/j.icarus.2019.113507

Battalio, M., & Wang, H. (2021). The Mars Dust Activity Database (MDAD): A comprehensive statistical study of dust storm sequences. *Icarus, 354*, 114059. https://doi.org/10.1016/j.icarus.2020.114059

Battalio, J. M., Martínez, G., Newman, C., de la Torre Juarez, M., Sánchez-Lavega, A., & Víudez-Moreiras, D. (2022). Planetary waves traveling between Mars Science Laboratory and Mars 2020. *Geophysical Research Letters, 49,* e2022GL100866. https://doi.org/10.1029/2022GL100866

Bell lll, J. F., Wolff, M. J., Malin, M. C., Calvin, W. M., Cantor, B. A., Caplinger, M. A., et al. (2009). Mars Reconnaissance Orbiter Mars Color Imager (MARCI): Instrument description, calibration, and performance. *Journal of Geophysical Research: Planets, 114*(E8), 1-41. https://doi.org/10.1029/2008je003315

Cantor, B., Malin, M., & Edgett, K. S. (2002). Multiyear Mars Orbiter Camera (MOC) observations of repeated Martian weather phenomena during the northern summer season. *Journal of Geophysical Research, 107*(E3), 5014. https://doi.org/10.1029/2001JE001588

Cantor, B. A., James, P. B., & Calvin, W. M. (2010). MARCI and MOC observations of the atmosphere and surface cap in the north polar region of Mars. *Icarus, 208*(1), 61–81. https://doi.org/10.1016/j.icarus.2010.01.032

Chen-Chen, H., Pérez-Hoyos, S., & Sánchez-Lavega, A. (2019). Characterisation of Martian dust aerosol phase function from sky radiance measurements by MSL engineering cameras. *Icarus, 330*, 16–29. https://doi.org/10.1016/j.icarus.2019.04.004

Clancy, R.T., Montmessin, F., Benson, J., Daerden, F., Colaprete, A., Wolff, M.J., Haberle, R.M., Forget, F., Smith, M.D., Zurek, R.W. (2017). Mars Clouds. In R. M. Haberle, R. Clancy, F. Forget, M. D. Smith, & R. W. Zurek (Eds.), *The atmosphere and climate of Mars* (pp. 76-105). Cambridge University Press.  https://doi.org/10.1017/9781139060172.005

Collins, M., Lewis, S. R., Read, P. L., & Hourdin, F. (1996). Baroclinic wave transitions in the martian atmosphere. *Icarus*, **120** (2), 344-357.

Forget, F., Hourdin, F., Fournier, R., Hourdin, C., Talagrand, O., Collins, M., Lewis, S. R., et al. (1999). Improved general circulation models of the Martian atmosphere from the surface to above 80 km. *Journal of Geophysical Research: Planets, 104*(E10), 24155–24175. https://doi.org/10.1029/1999je001025

Gebhardt, C., Guha, B. K., Young, R. M. B., Wolff, M. J., & Edwards, C. S. (2023). Sub-hourly observations of dust storm growth, lee waves, and Lyot Crater, by the EMM camera EXI. Geophysical Research Letters, 50, e2023GL105317. https://doi.org/10.1029/2023GL105317
Gierasch, P., Thomas, P., French, R., & Veverka, J. (1979). Spiral clouds on Mars: A new atmospheric phenomenon. *Geophysical Research Letters, 6*(5), 405–408. https://doi.org/10.1029/gl006i005p00405

Greybush, S. J., Gillespie, H. E., & Wilson, R. J. (2019). Transient eddies in the TES/MCS Ensemble Mars Atmosphere Reanalysis System (EMARS). *Icarus, 317*, 158–181. https://doi.org/10.1016/j.icarus.2018.07.001

Guha, B. K., Gebhardt, C., Young, R. M. B., Wolff, M. J., & Montabone, L. (2024). Seasonal and diurnal variations of dust storms in Martian year 36 based on the EMM-EXI database. Journal of Geophysical Research: Planets, 129, e2023JE008156. https://doi.org/10.1029/2023JE008156

Guzewich, S. D., Toigo, A. D., Kulowski, L., & Wang, H. (2015). Mars Orbiter Camera climatology of textured dust storms. *Icarus, 258*, 1–13. https://doi.org/10.1016/j.icarus.2015.06.023

Haberle, R. M., de la Torre Juarez, M., Kahre, M. A., Kass, D. M., Barnes, J. R., Hollingsworth, J. L., et al. (2018). Detection of northern hemisphere transient eddies at Gale crater Mars. *Icarus*, *307*, 150–160.  https://doi.org/10.1016/j.icarus.2018.02.013

Harri, A. -M., Genzer, M., Kemppinen, O., Kahanpää, H., Gomez-Elvira, J., Rodriguez-Manfredi, J. A., et al. (2014). Pressure observations by the Curiosity rover: Initial results. *Journal of Geophysical Research: Planets, 119*(1), 82–92.

https://doi.org/10.1002/2013je004423

Harri, A., Paton, M., Hieta, M., Polkko, J., Newman, C., Pla-Garcia, J., et al. (2024). Perseverance MEDA Atmospheric Pressure Observations - Initial Results. *Journal of Geophysical Research: Planets, 129*(3), e2023JE007880. https://doi.org/10.1029/2023je007880

Heavens, N. G. (2017). Textured Dust Storm Activity in Northeast Amazonis–Southwest Arcadia, Mars: Phenomenology and Dynamical Interpretation. *Journal of the Atmospheric Sciences, 74*(4), 1011–1037. https://doi.org/10.1175/jas-d-16-0211.1

Hernández-Bernal, J., Sánchez-Lavega, A., del Río-Gaztelurrutia, T., Hueso, R., Cardesín-Moinelo, A., Ravanis, E. M., et al. (2019). The 2018 Martian Global Dust Storm Over the South Polar Region Studied With MEx/VMC. *Geophysical Research Letters, 46*(17–18), 10330–10337. https://doi.org/10.1029/2019gl084266

Hernández-Bernal, J., Sánchez-Lavega, A., del Río-Gaztelurrutia, T., Ravanis, E., Cardesín-Moinelo, A., Connour, K., et al. (2021). An Extremely Elongated Cloud Over Arsia Mons Volcano on Mars: I. Life Cycle. *Journal of Geophysical Research: Planets, 126*(3), e2020JE006517. https://doi.org/10.1029/2020je006517

Hernández-Bernal, J., Spiga, A., Forget, F., & Banfield, D. (2024a). High order harmonics of thermal tides observed in the atmosphere of Mars by the pressure sensor on the InSight lander. *Geophysical Research Letters*, 51, e2023GL107674. https://doi.org/10.1029/2023GL107674

Hernández-Bernal J., Alejandro Cardesin-Moinelo, A., Hueso, R., Ravanis, E., Burgos-Sierra, A., Simon Wood., et al. (submitted 2024b). The Visual Monitoring Camera (VMC) on Mars Express: a new science instrument made from an old webcam orbiting Mars. *Planetary and Space Science*, 251, 105972 (2024)
 https://doi.org/10.1016/j.pss.2024.105972

Hess, S. L., Henry, R. M., Leovy, C. B., Ryan, J. A., & Tillman, J. E. (1977). Meteorological results from the surface of Mars: Viking 1 and 2. *Journal of Geophysical Research*, *82*(28), 4559–4574. https://doi.org/10.1029/js082i028p04559

Hess, S. L., Ryan, J. A., Tillman, J. E., Henry, R. M., & Leovy, C. B. (1980). The annual cycle of pressure on Mars measured by Viking landers 1 and 2. *Geophysical Research Letters*, *7*(3), 197–200. https://doi.org/10.1029/gl007i003p00197

Hinson, D. P., & Wilson, R. J. (2004). Temperature inversions, thermal tides, and water ice clouds in the Martian tropics. *Journal of Geophysical Research: Planets, 109*(1), E01002. https://doi.org/10.1029/2003je002129

Hinson, D. P., & Wang, H. (2010). Further observations of regional dust storms and baroclinic eddies in the northern hemisphere of Mars. *Icarus, 206*(1), 290–305. https://doi.org/10.1016/j.icarus.2009.08.019

Hinson, D., & Wilson, R. J. (2021). Baroclinic waves in the northern hemisphere of Mars as observed by the MRO Mars climate sounder and the MGS thermal emission spectrometer. *Icarus*, *357*, 114152. https://doi.org/10.1016/j.icarus.2020.114152

Hollingsworth J.L. and Kahre, M.A. (2014) Modeling Mars cyclogenesis and frontal waves: seasonal variations and implications on dust activity, Mars Atmospheric Modelling and Observations, 5th International Workshop, Oxford (UK), 20140011424. https://www-mars.lmd.jussieu.fr/oxford2014/abstracts/hollingsworth_oxford2014.pdf

Hunt, G. E., & James, P. B. (1979). Martian extratropical cyclones. *Nature, 278*(5704), 531–532. https://doi.org/10.1038/278531a0

ISIS-USGS (2024), [Software] https://isis.astrogeology.usgs.gov/7.0.0/index.html, https://github.com/DOI-USGS/ISIS3

Jaakonaho, I., Hieta, M., Genzer, M., Polkko, J., Mäkinen, T., Sánchez-Lavega, A., et al. (2023). Pressure sensor for the Mars 2020 Perseverance rover. *Planetary and Space Science, 239*, 105815. https://doi.org/10.1016/j.pss.2023.105815

James, I.N., Gray, L.J., 1986. Concerning the effect of surface drag on the circulation of a baroclinic planetary atmosphere. *Q. J. Roy. Meteor. Soc.* 112, 1231–1250. http://dx.doi.org/10.1002/qj.49711247417.

James, P. B., Christensen, P. R., Clancy, R. T., Lemmon, M. T., Withers, P. (2017). History of Mars atmosphere observations. In R. M. Haberle, R. Clancy, F. Forget, M. D. Smith, & R. W. Zurek (Eds.), *The atmosphere and climate of Mars* (pp. 20-41). Cambridge University Press. https://doi.org/10.1017/9781139060172.003

Kahre, M. A., Murphy, J. R., Newman, C. E., Wilson, R. J., Cantor, B. A., Lemmon, M. T., Wolff, M. J. (2017). The Mars dust cycle. In R. M. Haberle, R. Clancy, F. Forget, M. D. Smith, & R. W. Zurek (Eds.), *The atmosphere and climate of Mars* (pp. 295-337). Cambridge University Press. https://doi.org/10.1017/9781139060172.010

Kass, D. M., Kleinböhl, A., McCleese, D. J., Schofield, J. T., & Smith, M. D. (2016). Interannual similarity in the Martian atmosphere during the dust storm season. *Geophysical Research Letters, 43*(12), 6111–6118. https://doi.org/10.1002/2016gl068978

Lembo, V., Bordi, I., & Speranza, A. (2017). Annual and semiannual cycles of midlatitude near-surface temperature and tropospheric baroclinicity: reanalysis data and AOGCM simulations. *Earth System Dynamics, 8*(2), 295–312. https://doi.org/10.5194/esd-8-295-2017

Lemmon, M. T., Wolff, M. J., Bell III, J. F., Smith, M. D., Cantor, B. A., & Smith, P. H. (2015). Dust aerosol, clouds, and the atmospheric optical depth record over 5 Mars years of the Mars Exploration Rover mission. *Icarus, 251*, 96–111.

https://doi.org/10.1016/j.icarus.2014.03.029

Lemmon, M. T., Smith, M. D., Viudez-Moreiras, D., de la Torre-Juarez, M., Vicente-Retortillo, A., Munguira, A., et al. (2022). Dust, Sand, and Winds Within an Active Martian Storm in Jezero Crater. *Geophysical Research Letters, 49*(17), e2022GL100126. https://doi.org/10.1029/2022gl100126

Lemmon, M. T., Guzewich, S. D., Battalio, J. M., Malin, M. C., Vicente-Retortillo, A., Zorzano, M.-P., et al. (2024). The Mars Science Laboratory record of optical depth measurements via solar imaging. *Icarus, 408*, 115821. https://doi.org/10.1016/j.icarus.2023.115821

Leovy, C., and Zurek, R. Thermal tides and martian dust storms: Direct evidence for coupling. *J. Geophys. Res.* **84**, 2956-2968 (1979).

Lewis, S. R., Collins, M., Read, P. L., Forget, F., Hourdin, F., Fournier, R., et al. (1999). A climate database for Mars. *Journal of Geophysical Research, 104*(E10), 24177–24194. https://doi.org/10.1029/1999je001024

Lindzen, R. S., & Farrell, B. (1980). A Simple Approximate Result for the Maximum Growth Rate of Baroclinic Instabilities. *Journal of the Atmospheric Sciences, 37*(7), 1648–1654. https://doi.org/10.1175/1520-0469(1980)037<1648:asarft>2.0.co;2

Martinez, G. M., Newman, C. N., De Vicente-Retortillo, A., Fischer, E., Renno, N. O., Richardson, M. I., et al. (2017). The modern nearsurface Martian climate: A review of in-situ meteorological data from Viking to Curiosity. *Space Science Reviews*, *212*(1), 295–338. https://doi.org/10.1007/s11214-017-0360-x

MARCI images (2024), [Dataset] https://planetarydata.jpl.nasa.gov/img/data/mro/mars_reconnaissance_orbiter/marci/

Martín-Rubio C., Vicente-Retortillo A., Gómez F., Rodríguez-Manfredi J.A. (2024). Interannual variability of regional dust storms between Mars years 24 and 36. *Icarus*, 412, 115982. https://doi.org/10.1016/j.icarus.2024.115982

Millour, E., Forget, F., Spiga, A., Pierron, T., Bierjon, A., Montabone, L., et al. (2022). The Mars Climate Database (Version 6.1). In *Europlanet Science Congress 2022*, (EPSC2022-786). https://doi.org/10.5194/epsc2022-786

Montabone, L., Forget, F., Millour, E., Wilson, R. J., Lewis, S. R., Cantor, B., et al. (2015). Eight-year climatology of dust optical depth on Mars. *Icarus, 251*, 65–95. https://doi.org/10.1016/j.icarus.2014.12.034

Mulholland, D. P., Lewis, S. R., Read, P. L., Madeleine, J.-B., Forget, F. (2016). The solsticial pause on Mars: 2 modelling and investigation of causes, *Icarus*, 264, 465-477. http://dx.doi.org/10.1016/j.icarus.2015.08.038

Munguira, A., Hueso, R., Sánchez-Lavega, A., de la Torre-Juarez, M., Martínez, G. M., Newman, C. E., Sebastian, E., et al. (2023). Near Surface Atmospheric Temperatures at Jezero From Mars 2020 MEDA Measurements. *Journal of Geophysical Research: Planets, 128*(3), e2022JE007559. https://doi.org/10.1029/2022je007559

Montabone, L., Kleinboehl, A., Smith, M., Edwards, C., Forget, F., Kass, D., Millour, E., Stcherbinine, A. (2023). Reconstructing Martian Year 36 column dust optical depth maps using EMM/EMIRS and MRO/MCS. EGU General Assembly Conference Abstracts, EGU23-10341.
https://doi.org/10.5194/egusphere-egu23-10341

Newman C.E., M. de la Torre Juárez, J. Pla-García, R.J. Wilson, S.R. Lewis, L. Neary, M.A. Kahre, F. Forget, A. Spiga, M.I. Richardson, F. Daerden, T. Bertrand, D. Viúdez-Moreiras, R. Sullivan, A. Sánchez-Lavega, B. Chide, J.A. Rodriguez-Manfredi, (2021). Multi-model Meteorological and Aeolian Predictions for Mars 2020 and the Jezero Crater Region. *Space Science Review* 217:20. https://doi.org/10.1007/s11214-020-00788-2

Ordóñez-Etxeberria I., Sánchez-Lavega, A., Ordorika, T., Hueso, R. (2022), MeteoMars, a Tool to Explore Meteorological Events on Mars. In *Seventh International Workshop on the Mars Atmosphere: Modelling and Observations, p. 1508*.
https://www-mars.lmd.jussieu.fr/paris2022/abstracts/poster_Ordonez-Etxeberria_Inaki.pdf

Ormston, T., Denis, M., Scuka, D., & Griebel, H. (2011). An ordinary camera in an extraordinary location: Outreach with the Mars Webcam. *Acta Astronautica, 69*(7–8), 703–713. https://doi.org/10.1016/j.actaastro.2011.04.015

Patel, P., Tamppari, L., de la Torre Juárez, M., Lemmon, M., Coates, A., Wolff, M., et al. (2023). Geometric Properties of Water-ice Clouds as Observed from Jezero Crater in the First 600 sols with the NavCam Instrument On Board the Mars2020 Rover, Perseverance. *The Planetary Science Journal, 4*(12), 226. https://doi.org/10.3847/psj/acfc35

QGIS (2024), [Software], https://www.qgis.org/en/site/forusers/download.html

Rodriguez-Manfredi, J. A., de la Torre Juárez, M., Alonso, A., Apéstigue, V., Arruego, I., Atienza, T., et al. (2021). The Mars Environmental Dynamics Analyzer, MEDA. A Suite of Environmental Sensors for the Mars 2020 Mission. *Space Science Reviews, 217*(3), 48. https://doi.org/10.1007/s11214-021-00816-9

Rodriguez-Manfredi, J. A., & de la Torre Juarez, M. (2021). The Mars environmental dynamics analyzer, MEDA. *NASA Planetary Dcata System* [Dataset]
https://doi.org/10.17189/1522849

Rodriguez-Manfredi, J. A., de la Torre Juarez, M., Sanchez-Lavega, A., Hueso, R., Martinez, G., Lemmon, M. T., et al. (2023). The diverse meteorology of Jezero crater over the first 250 sols of Perseverance on Mars. *Nature Geoscience, 16*(1), 19–28. https://doi.org/10.1038/s41561-022-01084-0

Ryan, J., & Henry, R. (1979). Mars atmospheric phenomena during major dust storms, as measured at surface. *Journal of Geophysical Research*, 84(B6), 2821–2829. https://doi.org/10.1029/jb084ib06p02821.

Sánchez-Lavega, A., Chen-Chen, H., Ordoñez-Etxeberria, I., Hueso, R., del Río-Gaztelurrutia, T., Garro, A., et al. (2018a). Limb clouds and dust on Mars from images obtained by the Visual Monitoring Camera (VMC) onboard Mars Express. *Icarus, 299*, 194–205. https://doi.org/10.1016/j.icarus.2017.07.026

Sánchez-Lavega, A., Garro, A., del Río-Gaztelurrutia, T., Hueso, R., Ordoñez-Etxeberria, I., Chen Chen, H., et al. (2018b). A Seasonally Recurrent Annular Cyclone in Mars Northern Latitudes and Observations of a Companion Vortex. *Journal of Geophysical Research: Planets, 123*(11), 3020–3034. https://doi.org/10.1029/2018je005740

Sánchez-Lavega, A., Erkoreka, A., Hernández-Bernal, J., del Río-Gaztelurrutia, T., García-Morales, J., Ordoñez-Etxeberría, I., et al. (2022). Cellular patterns and dry convection in textured dust storms at the edge of Mars North Polar Cap. *Icarus, 387*, 115183. https://doi.org/10.1016/j.icarus.2022.115183

Sánchez-Lavega, A., del Rio-Gaztelurrutia, T., Hueso, R., Juárez, M. de la T., Martínez, G. M., Harri, A. -M., et al. (2023). Mars 2020 Perseverance Rover Studies of the Martian Atmosphere Over Jezero From Pressure Measurements. *Journal of Geophysical Research: Planets, 128*(1), e2022JE007480. https://doi.org/10.1029/2022je007480

Sánchez-Lavega, A., del Río-Gaztelurrutia, T., Spiga, A., Hernández-Bernal, J., Larsen, E., Tirsch, D., et al. (2024). Dynamical Phenomena in the Martian Atmosphere Through Mars Express Imaging. *Space Science Reviews, 220*(1): 16. https://doi.org/10.1007/s11214-024-01047-4

Smith, M. D., Martínez, G. M., Sebastián, E., Lemmon, M. T., Wolff, M. J., Apéstigue, V., et al. (2023). Diurnal and seasonal variations of aerosol optical depth observed by MEDA/TIRS at Jezero Crater, Mars. *Journal of Geophysical Research: Planets*, 128, e2022JE007560. https://doi.org/10.1029/2022JE007560

Steele, L. J., Kleinböhl, A., Kass, D. M., & Zurek, R. W. (2021). Aerosols and Tides in the Martian Tropics During Southern Hemisphere Spring Equinox From Mars Climate Sounder Data. *Journal of Geophysical Research: Planets, 126*(4), e2020JE006776. https://doi.org/10.1029/2020je006776

Toledo, D., Gomez, L., Apestigue, V., Arruego, I., Smith, M., Munguira, A., et al. (2023). Twilight mesospheric clouds in Jezero as observed by MEDA Radiation and Dust Sensor (RDS). *Journal of Geophysical Research:Planets*, *128*(7), e2023JE007785. https://doi.org/10.1029/2023JE007785

VMC images (2024), [Dataset] https://psa.esa.int/psa/#/pages/home

Wang, H., & Ingersoll, A. P. (2002). Martian clouds observed by Mars Global Surveyor Mars Orbiter Camera. *Journal of Geophysical Research: Planets, 107*(E10), 5078. https://doi.org/10.1029/2001je001815

Wilson, R. J. and Hamilton, K. (1996). Comprehensive model simulation of thermal tides in the Martian atmosphere. J. Atmos. Sci. 53, 1290-1326. https://doi.org/10.1175/1520-0469(1996)053<1290:CMSOTT>2.0.CO;2

Wilson, R. J., Lewis, S. R., & Montabone, L. (2008). Thermal tides in an assimilation of three years of thermal emission spectrometer data from Mars Global Surveyor. In *Third international workshop on the Mars atmosphere: Modelling and observations workshop*. https://www.lpi.usra.edu/meetings/modeling2008/pdf/9022.pdf

Wolfe, C. A., Edwards, C. S., Smith, M. D., & Christensen, P. R. (2023). Constraining changes in surface dust thickness on Mars using diurnal surface temperature observations from EMIRS. Journal of Geophysical Research: Planets, 128, e2023JE007794. https://doi.org/10.1029/2023JE007794

Wolff, M. J., Clancy, R. T., Kahre, M. A., Haberle, R. M., Forget, F., Cantor, B. A., & Malin, M. C. (2019). Mapping water ice clouds on Mars with MRO/MARCI. *Icarus, 332*, 24–49. https://doi.org/10.1016/j.icarus.2019.05.041

Zurek, R. W., & Smrekar, S. E. (2007). An overview of the Mars Reconnaissance Orbiter (MRO) science mission. *Journal of Geophysical Research: Planets, 12*(E5), E05S01. https://doi.org/10.1029/2006je002701

Zurek, R. W., (2017). Understanding Mars and Its Atmosphere. In R. M. Haberle, R. Clancy, F. Forget, M. D. Smith, & R. W. Zurek (Eds.), *The atmosphere and climate of Mars* (pp. 3-19). Cambridge University Press. https://doi.org/10.1017/9781139060172.002

Zurita-Zurita, S., de la Torre Juárez, M., Newman, C. E., Viúdez-Moreiras, D., Kahanpää, H. T., Harri, A. -M., et al. (2022). Mars Surface Pressure Oscillations as Precursors of Large Dust Storms Reaching Gale. *Journal of Geophysical Research: Planets, 127*(8), e2021JE007005. https://doi.org/10.1029/2021je007005